\documentclass[aps,prb,longbibliography,reprint,showpacs,superscriptaddress,amsmath]{revtex4-1}
\usepackage{color}
\usepackage{graphicx}
\usepackage{amsmath}
\usepackage{comment}

\DeclareMathOperator\arctanh{arctanh}

\begin{document}

\title{Evolution of nematic fluctuations in CaK(Fe$_{1-x}$Ni$_x$)$_4$As$_4$ with spin-vortex crystal magnetic order} 

\author{Anna E. B\"ohmer}
\affiliation{Institute for Quantum Materials and Technologies, Karlsruhe Institute of Technology, 76021 Karlsruhe, Germany}
\affiliation{Ames Laboratory, U.S. DOE, Iowa State University, Ames, Iowa 50011, USA}

\author{Fei Chen}
\affiliation{School of Physics and Astronomy, University of Minnesota, Minneapolis, MN 55455, USA}

\author{William R. Meier}
\affiliation{Ames Laboratory, U.S. DOE, Iowa State University, Ames, Iowa 50011, USA}
\affiliation{Department of Physics and Astronomy, Iowa State University, Ames, Iowa 50011, USA}
\affiliation{Materials Science \& Technology Division, Oak Ridge National Laboratory, Oak Ridge, Tennessee 37831, USA}

\author{Mingyu Xu}
\affiliation{Ames Laboratory, U.S. DOE, Iowa State University, Ames, Iowa 50011, USA}
\affiliation{Department of Physics and Astronomy, Iowa State University, Ames, Iowa 50011, USA}

\author{Gil Drachuck}
\affiliation{Ames Laboratory, U.S. DOE, Iowa State University, Ames, Iowa 50011, USA}
\affiliation{Department of Physics and Astronomy, Iowa State University, Ames, Iowa 50011, USA}

\author{Michael Merz}
\affiliation{Institute for Quantum Materials and Technologies, Karlsruhe Institute of Technology, 76021 Karlsruhe, Germany}

\author{Paul Wiecki}
\affiliation{Institute for Quantum Materials and Technologies, Karlsruhe Institute of Technology, 76021 Karlsruhe, Germany}

\author{Sergey L. Bud'ko}
\affiliation{Ames Laboratory, U.S. DOE, Iowa State University, Ames, Iowa 50011, USA}
\affiliation{Department of Physics and Astronomy, Iowa State University, Ames, Iowa 50011, USA}

\author{Vladislav Borisov}
\affiliation{Department of Physics and Astronomy, Uppsala University, Box 516, Uppsala, SE-75120, Sweden}

\author{Roser Valent\'i}
\affiliation{Institut f\"ur Theoretische Physik, Goethe-Universit\"at Frankfurt, Max-von-Laue-Strasse 1, 60438 Frankfurt am Main, Germany}

\author{Morten H. Christensen}
\affiliation{School of Physics and Astronomy, University of Minnesota, Minneapolis, MN 55455, USA}

\author{Rafael M. Fernandes}
\affiliation{School of Physics and Astronomy, University of Minnesota, Minneapolis, MN 55455, USA}

\author{Christoph Meingast}
\affiliation{Institute for Quantum Materials and Technologies, Karlsruhe Institute of Technology, 76021 Karlsruhe, Germany}

\author{Paul C. Canfield}
\affiliation{Ames Laboratory, U.S. DOE, Iowa State University, Ames, Iowa 50011, USA}
\affiliation{Department of Physics and Astronomy, Iowa State University, Ames, Iowa 50011, USA}

\date{\today}
             
\begin{abstract}
 The CaK(Fe$_{1-x}$Ni$_x$)$_4$As$_4$ superconductors resemble the archetypal 122-type iron-based materials but have a crystal structure with distinctly lower symmetry. This family hosts one of the few examples of the so-called spin-vortex crystal magnetic order, a non-collinear magnetic configuration that preserves tetragonal symmetry, in contrast to the orthorhombic collinear stripe-type magnetic configuration common to the 122-type systems. Thus, nematic order is completely absent from its phase diagram. To investigate the evolution of nematic fluctuations in such a case, we present elastoresistance and elastic modulus measurements in CaK(Fe$_{1-x}$Ni$_x$)$_4$As$_4$ ($x=0-0.05$) combined with phenomenological modeling and density functional theory. We find clear experimental signatures of considerable nematic fluctuations, including softening of the Young's modulus $Y_{[110]}$ and a Curie-Weiss type divergence of the $B_{2g}$ elastoresistance coefficient in CaK(Fe$_{0.951}$Ni$_{0.049}$)$_4$As$_4$. Overall, nematic fluctuations within this series bear strong similarities to the hole-doped Ba$_{1-x}$K$_x$Fe$_2$As$_2$ series, including a substitution-induced sign change. Our theoretical analysis addresses the effect of the specific crystal symmetry of the 1144-type structure in determining its magnetic ground state and on the nematic fluctuations. 
\end{abstract}

\maketitle
\section{Introduction}
Magnetism in the iron-based superconductors exhibits an intriguing complexity. The most common magnetic order is the stripe spin-density wave (SSDW) phase.\cite{Cruz2008,Dai15_review} In the SSDW phase two separate symmetries are broken, namely the spin-rotational symmetry and the lattice rotational symmetry. Crucially, fluctuations of the magnetic order parameter can break the lattice rotational symmetry prior to the onset of magnetic order\cite{Chandra1990}. This results in a nematic phase which can be seen as a vestigial phase of the underlying SSDW order\cite{Fernandes2019}. As nematic order breaks the four-fold rotational symmetry of the tetragonal crystal, a tetragonal-to-orthorhombic structural transition occurs at the nematic transition\cite{Nandi2010}. 
Chemical substitution suppresses both magnetic and nematic transition temperatures and gives rise to superconductivity in a large number of materials\cite{Johnston2010}. Intriguingly, the highest superconducting transition temperature occurs near the substitution level where not only the magnetic, but also the nematic transition temperature extrapolates to zero, suggesting a nematic quantum critical point\cite{Worasaran20}. This has motivated proposals that nematic fluctuations play a key role in superconducting pairing\cite{Lederer2015,Schattner2016}. 
However, different magnetic orders that preserve the tetragonal crystal symmetry have been observed as well\cite{Kim2010_III,Avci2014,Boehmer2015II,Wang2016II,Allred2016,Taddei2017,Meier2018,Wang2018,Sheveleva2020} and, in some compounds, long-range nematic order does not persist down to zero temperature. This raises the question of how the nematic fluctuations evolve when the magnetic ground state preserves the tetragonal crystal symmetry.

Generally, magnetic order in iron-based compounds can be described by considering the spatial variation of the Fe moments $\mathbf{m}$ at positions $\mathbf{R}$, $\mathbf{m}(\mathbf{R})=\mathbf{M}_1\cos(\mathbf{Q}_1\cdot\mathbf{R})+\mathbf{M}_2\cos(\mathbf{Q}_2\cdot\mathbf{R})$. There are two magnetic order parameters, $\mathbf{M}_{1,2}$, describing stripes along two orthogonal and symmetry-equivalent directions $\mathbf{Q}_{1,2}$\cite{Lorenzana2008,Fernandes2016}. In this description, the aforementioned SSDW phase is given by $\mathbf{M}_1\neq 0$ and $\mathbf{M}_2=0$ (or vice versa). In many hole-doped systems $AE_{1-x}A_x$Fe$_2$As$_2$ ($AE$=Ba,Sr,Ca and $A$=K,Na) the SSDW phase yields to other magnetic phases as doping is increased \cite{Taddei2017}.  
In particular, the magnetic ground state changes from the typical SSDW to the so-called charge-spin density wave (CSDW) phase, characterized by $\mathbf{M}_1 \parallel \mathbf{M}_2$, upon increasing doping in many of these hole-doped 122's\cite{Allred2016,Taddei2017}. In the most detailed phase diagram of (Sr,Na)Fe$_2$As$_2$, several other ordered phases have been mapped out near optimal doping \cite{Wang2019}, the nature of some of which remains unclear \cite{Yi2018,Sheveleva2020}.
Notably, in (Sr,Na)Fe$_2$As$_2$, neutron scattering experiments have shown the persistence of nematic fluctuations even in the tetragonal CSDW phase \cite{Frandsen2018}. This persistence of nematic fluctuations may be a consequence of the proximity to the SSDW magnetic state present over a wide range of compositions in the underdoped region of the phase diagram.

Here, we address the issue of nematic fluctuations in systems with tetragonal magnetic states through elastoresistivity and elastic modulus measurements of the Ni-doped 1144-compound, CaK(Fe$_{1-x}$Ni$_x$)$_4$As$_4$. This is one of the few compounds for which both the SSDW magnetic order and nematic order are completely absent from the phase diagram\cite{Meier2018}. Stoichiometric CaKFe$_4$As$_4$ does not order magnetically and has a rather high superconducting transition temperature of $T_c=35$ K\cite{Iyo2016,Meier2016}. Magnetic order can, however, be stabilized by electron-doping, e.g., through partial replacement of Fe by Ni or Co\cite{Meier2018}.
This magnetic phase occurs in the tetragonal structure and is described by $|\mathbf{M}_1|=|\mathbf{M}_2|$ and $\mathbf{M}_1 \perp \mathbf{M}_2$\cite{Meier2018}. In this non-collinear configuration, the in-plane magnetic moments wind either clockwise or counterclockwise as a plaquette of Fe atoms is traversed, which motivates the name spin-vortex crystal (SVC) phase \cite{Fernandes2016,OHalloran2017}. Just as the SSDW phase gives rise to a vestigial nematic order, so does the SVC phase give rise to a vestigial order of its own, namely a phase with finite spin-vorticity, $\mathbf{M}_1 \times \mathbf{M}_2$, termed spin-vorticity density wave (SVDW) \cite{Fernandes2016}. 

From a charge count perspective, CaKFe$_4$As$_4$ is analogous to 50\% hole-doped 122 systems, e.g. Ba$_{0.5}$K$_{0.5}$Fe$_2$As$_2$, and the two materials indeed exhibit many similar physical properties\cite{Meier2016}. Nevertheless, the doping-induced magnetic ground states are different. The 122-type system  
has SSDW or CSDW magnetic ground states. Evidence for an SVC phase in a 122-type material has only recently been reported in a small part of the phase diagram of (Ba,Na)Fe$_2$As$_2$ \cite{Sheveleva2020}. In contrast, the SVC phase is the only magnetic order present in the phase diagram of Co and Ni substituted CaKFe$_4$As$_4$. 

\begin{figure}
\includegraphics[width=\columnwidth]{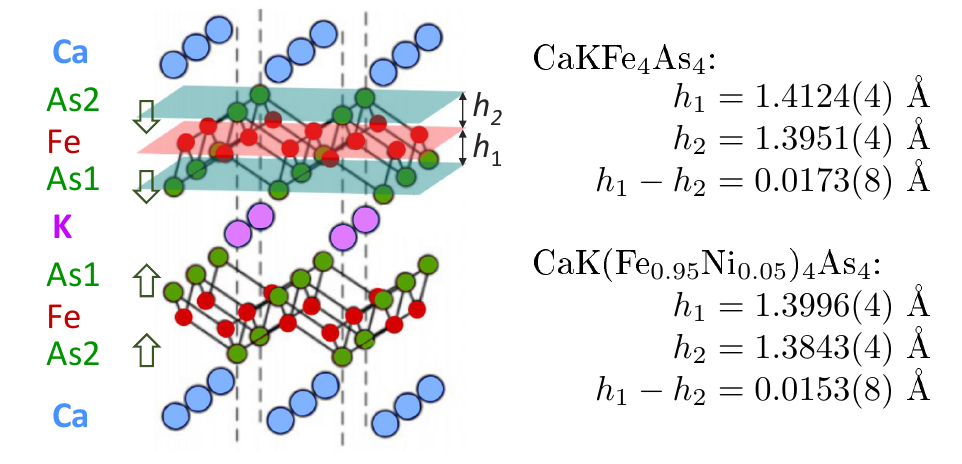}
\caption{\label{fig:crystal_structure}The 1144-type crystal structure with $P4/mmm$ space group. The green arrows indicate characteristic shifts of the As1 and As2 layers with respect to the iron plane that distinguishes this structure from 122-structure with space group $I4/mmm$. The height of the As1 and As2 layers above or below the Fe-plane, $h_1$ and $h_2<h_1$, are indicated.}
\end{figure}

The crucial ingredient for the stabilization of the SVC phase in doped CaKFe$_4$As$_4$ is likely the lower symmetry of its crystal structure\cite{Meierthesis}. In contrast to the 122 compounds, which crystallize in the $I4/mmm$ space group, these 1144 compounds lack a glide-plane symmetry and their space group is instead $P4/mmm$. The glide-plane symmetry is broken because the alkali ion, K, and the alkaline earth ion, Ca, form alternating layers in CaKFe$_4$As$_4$ (see Fig.~\ref{fig:crystal_structure}). As a consequence, there are two distinct As-sites, which has, e.g., been confirmed by NMR\cite{Cui2017}. Their inequivalence results in a symmetry-breaking field, $\eta$, which is oriented along the $c$-axis and has the same symmetry of the $z$-component of the spin-vorticity density wave order parameter (SVDW OP)\cite{Christensen2019}, $(\mathbf{M}_1 \times \mathbf{M}_2)\cdot \hat{z}$. Hence, the 1144 crystal structure naturally biases the system towards the SVC phase by inducing its vestigial order at high temperatures via bilinear coupling between $\eta$ and the SVDW OP\cite{Meier2018}.

The lack of SSDW and nematic orders makes CaK(Fe,Ni)$_4$As$_4$ ideally suited for a study of the impact of nematic fluctuations in systems exhibiting no long-range orthorhombic order of any kind. Furthermore, the impact of the field $\eta$, which vanishes by symmetry in the well-studied 122 compounds, on the nematic fluctuations is unclear. Here, we experimentally map out the nematic susceptibility of CaK(Fe$_{1-x}$Ni$_x$)$_4$As$_4$ ($x=0-0.05$) by means of elastoresistivity and elastic modulus measurements complemented by density functional theory. We find clear evidence for significant nematic fluctuations through the $B_{2g}$ nematic susceptibility. Our observations are consistent with theoretical calculations performed within a phenomenological Ginzburg-Landau framework, in which the presence of large nematic fluctuations is attributed to a close proximity between the SVC and SSDW instabilities in parameter space \cite{Christensen18}. 

After introducing the experimental and theoretical methods in Sec.~\ref{sec:methods}, we present results of single-crystal x-ray diffraction quantifying the structural asymmetry between the two As sites in Sec.~\ref{sec:diffraction}. We proceed to discuss how this inequivalence impacts the nematic susceptibility within a specific theoretical model in Sec.~\ref{sec:nemsus_theo}. Measurements of the elastoresistance of CaK(Fe$_{1-x}$Ni$_x$)$_4$As$_4$ in the $B_{1_g}$ and $B_{2_g}$ symmetry channels are presented in Sec.~\ref{sec:elastoresistance} whereas experimental and theoretical results for the elastic modulus are shown and discussed in Sec.~\ref{sec:elastic_modulus}.
Section~\ref{sec:conclusions} contains our conclusions.

\section{Methods\label{sec:methods}}
Single crystals of CaK(Fe$_{1-x}$Ni$_x$)$_4$As$_4$ ($x=0-0.05$) were grown out of a high-temperature solution rich in transition-metals and arsenic, as described in detail in Refs. \onlinecite{Meier2018,Meier2016,Meier2017}. The Ni content was determined via wavelength-dispersive x-ray spectroscopy as in Ref. \onlinecite{Meier2017}. Single-crystal x-ray diffraction (XRD) data were collected at 295 K on a STOE imaging plate diffraction system (IPDS-2T) using Mo $K_{\alpha}$ radiation. All accessible symmetry-equivalent reflections (about $5500$) were measured up to a maximum angle of $2 \Theta =65^\circ$. The data were corrected for Lorentz, polarization, extinction, and absorption effects. Around 195 averaged symmetry-independent reflections ($I > 2 \sigma$) were included for structure determination and for the corresponding refinements with SHELXL\cite{Scheldrick2008} and JANA2006\cite{Jana2006} in space group $P4/mmm$. 

For the elastoresistance measurements, we used pairs of thin ($\sim20$ $\mu$m) bar-like samples cut from one single crystal, oriented along the [110] (Fe-Fe bond direction) and the [100] crystallographic directions. Then, two samples for each of these directions were glued perpendicular to each other onto a piezoelectric stack (Piezomechanik) using Devcon 5-Minute Epoxy, see the insets in Fig.  \ref{fig:anistropic_resistance} for an illustration of the experimental configuration. The samples' resistances were measured in a 4-contact geometry, using a Lakeshore 370 or Lakeshore 372 resistance bridge and contacts were made with Dupont silver paint. The strain was measured via the resistance change of two perpendicular strain gauges attached to the opposite side of the piezostack. A Janis 950 closed cycle refrigerator with He exchange gas provided the temperature environment. 

The elastic Young's modulus (the elastic modulus for uniaxial deformation) was determined using a three-point bending method in a high-resolution capacitance dilatometer, as described in Ref. \onlinecite{Boehmer2014}. Complementary thermal expansivity measured in the same dilatometer\cite{Meingast1990} is also reported. 

 Theoretical modeling was performed within the Ginzburg-Landau framework and is described in detail in Appendix \ref{app:theory_details}. Bare elastic constants, unrenormalized by coupling to nematic fluctuations, were calculated using density functional theory while taking into account the spin-vortex nature of magnetism in this system, see Appendix \ref{app:DFT_details} for further details.

\begin{table}
\caption{\label{tab:refinement} Crystallographic data for CaKFe$_{4}$As$_{4}$ and CaK(Fe,Ni)$_{4}$As$_{4}$ at 295 K determined from single-crystal XRD. The structure was refined in the tetragonal space group $P4/mmm$\cite{Iyo2016}.
The $U_{ii}$ denote the anisotropic atomic displacement parameters (for all atoms $U_{12}=U_{13}=U_{23}=0$ and for Ca, K, As1, and As2 $U_{11}=U_{22}$). Fe and Ni were restrained to common $z$ and $U_{ii}$ parameters. The height $h_1$ ($h_2$) of the As1 (As2) layer above or below the (Fe,Ni) layer is also given.}
\begin{ruledtabular}
        \begin{tabular}{rrrccc}
      &  &Wyckoff position&$x$&$y$&$z$\\
      \hline
      &     Ca&1a&$0$&$0$ &$0$  \\
     &     K&1d&$\frac{1}{2}$&$\frac{1}{2}$ &$\frac{1}{2}$  \\
     &     (Fe,Ni)&4i&$\frac{1}{2}$&$0$ &$z_\textnormal{(Fe,Ni)}$  \\
     &     As1&2g&$0$&$0$ &$z_\textnormal{As1}$  \\
     &     As2&2h&$\frac{1}{2}$&$\frac{1}{2}$ &$z_{\textnormal{As2}}$  \\
        \end{tabular}
\end{ruledtabular}
\rule{0pt}{4ex}  
\begin{ruledtabular}
		\begin{tabular}[b]{rrccc}
		
    &    &  &  CaKFe$_{4}$As$_{4}$ & CaK(Fe,Ni)$_{4}$As$_{4}$ \\
			\hline
	&		&  $a$ (\r{A}) & 3.8710(5) & 3.8673(5) \\
	&		&  $c$ (\r{A}) & 12.8923(27) & 12.8142(25)  \\[1mm]
	& Ca		& $U_{11}$ (\r{A}$^2$) & 0.0142(20) & 0.0190(9)  \\
	&		& $U_{33}$ (\r{A}$^2$) & 0.0253(37) & 0.0353(16)   \\[1mm]
	& K		& $U_{11}$ (\r{A}$^2$) & 0.0254(26) &  0.0251(11)  \\
	&		& $U_{33}$ (\r{A}$^2$) & 0.0249(42) &  0.0378(18) \\[1mm]
	& (Fe,Ni) & $z$ & 0.23205(19) & 0.23165(9)   \\
	&      	& $U_{11}$ (\r{A}$^2$) & 0.0093(10) &  0.0159(6)  \\
	&      	& $U_{22}$ (\r{A}$^2$) & 0.0168(11) & 0.0158(5)  \\
	&		& $U_{33}$ (\r{A}$^2$) & 0.0255(12) & 0.0327(7)  \\[1mm]
	& As1    	& $z$ & 0.34159(18) &  0.34087(8)  \\
	&		& $U_{11}$ (\r{A}$^2$) & 0.0146(8) & 0.0182(4)  \\
	&		& $U_{33}$ (\r{A}$^2$) & 0.0284(16) & 0.0319(6)  \\
	&     	& $h_1$ (\r{A}) & 1.4124(4) &  1.3996(4)  \\[1mm]	
	& As2    	& $z$ & 0.12383(19) & 0.12362(8)   \\
	&		& $U_{11}$ (\r{A}$^2$) & 0.0157(9) & 0.0178(4)  \\
	&		& $U_{33}$ (\r{A}$^2$) & 0.0250(14) &  0.0312(6)  \\
	&     	& $h_2$ (\r{A}) & 1.3951(4) &  1.3843(4)  \\[1mm]
	&		& $R_1$ (\%) & 3.47 &  2.35 \\
	&		& $wR_2$ (\%) & 5.69 & 4.90 \\
	&		& GOF (\%) & 1.41 &  1.79 \\
			\end{tabular}
\end{ruledtabular}
\end{table}

\section{Structural refinement\label{sec:diffraction}}
The structural feature that singles out the 1144 materials is the broken glide symmetry of the Fe-plane, resulting from the different height of the As sites above and below the plane. 
To fully determine the crystal structure and quantify this asymmetry, we have performed single-crystal structural refinement of pure and 5\% Ni substituted CaK(Fe$_{1-x}$Ni$_x$)$_4$As$_4$. The results are summarized in Table \ref{tab:refinement}. All refinements converged quite well and show very good reliability factors, GOF, $R_1$, and $wR_2$. Our results for pure CaKFe$_4$As$_4$ are consistent with a previous report\cite{Mou2016}, but have much smaller error bars. Note that the similar structure factors of Fe and Ni do not allow an independent refinement of their structural parameters. Therefore, Fe and Ni were restricted to have the same $z$ coordinate and the same anisotropic atomic displacement parameters.

Symmetry considerations and extinction rules strongly indicate that indeed CaKFe$_{4}$As$_{4}$ and CaK(Fe,Ni)$_{4}$As$_{4}$ share the space group type $P4/mmm$ and Ca and K indeed occupy different Wyckoff positions, in complete agreement with previous structural data\cite{Iyo2016,Mou2016}, and, e.g., NMR results that find two distinct As sites\cite{Cui2017}. Further support for ordered Ca and K layers comes from the significantly different Ca-As and K-As bond lengths ($\approx 3.16$ vs $3.41$ \r{A}) derived from our refinements. However, a certain degree of disorder of Ca and K on the corresponding $1a$ and $1b$ Wyckoff positions (see Table \ref{tab:refinement}) cannot be ruled out, due to comparable structure factors. 

Indeed, the As1-plane is at a greater distance from the (Fe,Ni)-plane than the As2-plane, see Fig. \ref{fig:crystal_structure}. The difference of these heights amounts to $h_{2}-h_{1}=0.0173(8)$ \AA\  for CaKFe$_4$As$_4$ and $h_{2}-h_{1}=0.0153(8)$ \AA\  for CaK(Fe$_{0.95}$Ni$_{0.05}$)$_4$As$_4$, i.e., only $1.1\%$-$1.2\%$ of the mean distance. There is only a minor doping variation of $h_2-h_1$. We will therefore assume that the variation of the field $\eta$ with doping is negligible in the range of compositions/parameters studied. Note that, while $\eta$ is proportional to $h_1-h_2$, its magnitude cannot be determined due to unknown coupling constants.

\section{Theoretical modeling of the nematic susceptibility\label{sec:nemsus_theo}}

\begin{figure}
    \centering
    \includegraphics[width=\columnwidth]{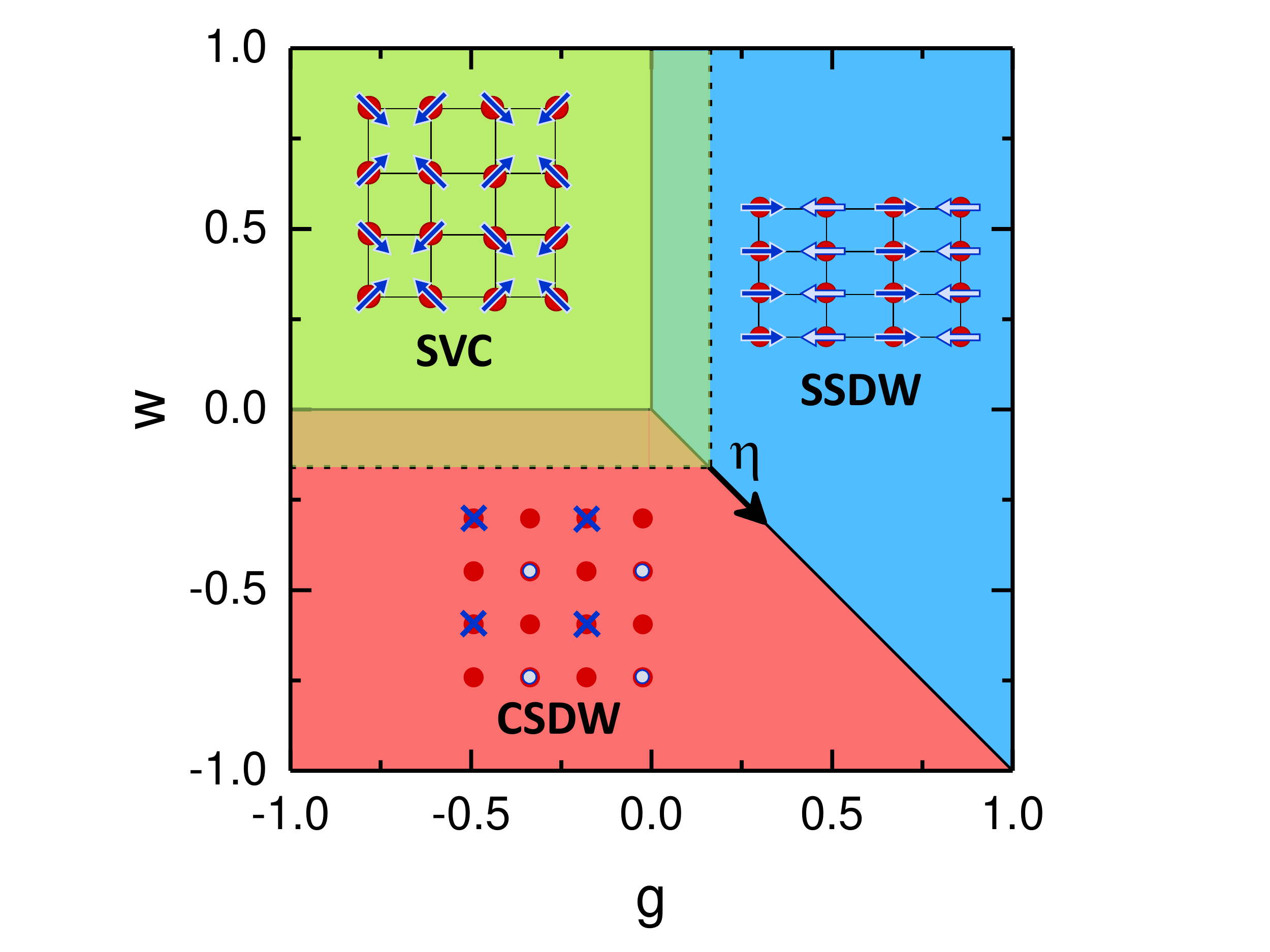}
    \caption{Mean-field phase diagram of the Ginzburg-Landau model of Eq.~\eqref{eq:free_energy} in the phase space of the Landau parameters $g$ and $w$, showing the enlargement of the SVC phase by the symmetry-breaking crystal field, $\eta$, related to inequivalent As1 and As2 sites. The magnetic moments in the spin-vortex crystal (SVC), stripe spin-density wave (SSDW) and charge-spin density wave (CSDW) phases are depicted in the Fe plane for $\eta=0$. Experimentally, they are in the plane in the SVC and SSDW phase, whereas the CSDW phase exhibits alternating magnetic moments perpendicular to the Fe-plane with every other Fe site bearing zero magnetic moment. The solid lines show the phase boundaries at $\eta=0$ and the dashed lines indicate the change due to a finite $\eta$.}
    \label{fig:phase_diagram}
\end{figure}

To better understand the different magnetic phases and the effect of the explicit broken symmetry between As1 and As2 on the nematic susceptibility, we performed a phenomenological analysis within the Ginzburg-Landau framework. We depart from the free energy 
\begin{align}
	\mathcal{F} &=  r_0 \left( \mathbf{M}_1^2 + \mathbf{M}_2^2 \right) + \frac{u}{2}  \left( \mathbf{M}_1^2 + \mathbf{M}_2^2 \right)^2 - \frac{g}{2} \left( \mathbf{M}_1^2 - \mathbf{M}_2^2 \right)^2 \nonumber \\ &+ 2 w  \left( \mathbf{M}_1 \cdot \mathbf{M}_2 \right)^2 - \eta \left(\mathbf{M}_1 \times \mathbf{M}_2 \right)\cdot \hat{z} \,, \label{eq:free_energy}
\end{align}
where $\mathbf{M}_{1,2}$ refers to the two magnetic order parameters discussed in the Introduction. The first four terms have been widely discussed previously in the context of, e.g. 122 iron pnictides (see for instance Refs. \onlinecite{Lorenzana2008,Wang2014,Fernandes2016}). The quadratic Ginzburg-Landau parameter is $r_0 \propto T - T_{\rm mag,0}$, where $T_{\rm mag,0}$ is a mean-field magnetic transition temperature. The quartic parameter $u$ ensures the stability of the functional whereas $g$ and $w$ determine the selected magnetic state (SSDW, SVC, or CSDW). Indeed, in the absence of $\eta$, the mean-field phase diagram is well-known~\cite{Wang2014}, see Fig.~\ref{fig:phase_diagram}. There are three different magnetic phases, and each is associated with a vestigial phase\cite{Fernandes2016,Christensen2019}. The magnetic phases, as discussed in the Introduction, are the SSDW phase with $\mathbf{M}_1 \neq 0$ and $\mathbf{M}_2 =0$ (or vice versa), the CSDW phase with $\mathbf{M}_1 \parallel \mathbf{M}_2$ and the SVC phase with $\mathbf{M}_1 \perp \mathbf{M}_2$. In the $g-w$ parameter space of Eq.~\eqref{eq:free_energy}, the phase boundaries are $\{ g<0, w=0 \}$, $\{ g=0, w>0 \}$, and $\{g>0, w<0, g=-w \}$.

What makes CaK(Fe,Ni)$_4$As$_4$ different is the inequivalence of the two As atoms (As1 and As2). As discussed in Ref. \onlinecite{Meier2018}, this gives rise to an effective crystal-field $\eta$ that acts as a conjugate field to the vestigial spin-vorticity density wave (SVDW) order parameter $(\mathbf{M}_1 \times \mathbf{M}_2)\cdot \hat{z}$, described by the last term in the free energy Eq.~\eqref{eq:free_energy}. We find that the mean-field phase diagram changes when $\eta$ is finite. The SVC phase expands with new boundaries located at $w=-\eta\frac{u}{2r_0+\eta}$ and $g=\eta\frac{u}{2r_0+\eta}$, while the line separating the SSDW and CSDW phases remains $g=-w$, see Fig.~\ref{fig:phase_diagram}. A finite $\eta$ also impacts the form of the SSDW and CSDW order parameters which acquire a small SVC component. This implies that one can change the ground state between the SSDW and SVC phase or between the CSDW and SVC phase by varying either $\eta$ or $r_0$. As a consequence, the phase boundaries in Fig.~\ref{fig:phase_diagram} now become functions of $r_0$ and $\eta$ and therefore move as a function of temperature, in this approximation.

\begin{figure*}
    \centering
    \includegraphics[width=17.2cm]{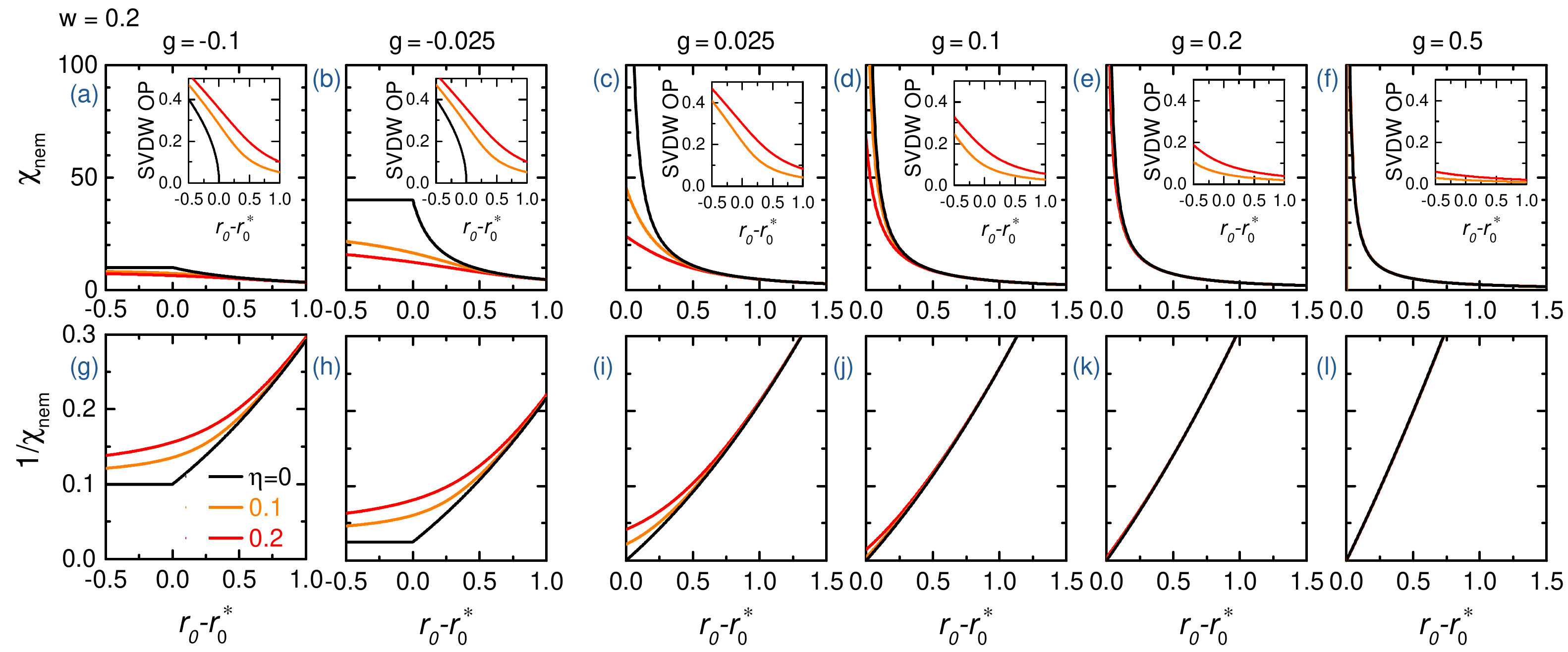}
    \caption{(a)-(f) The nematic susceptibility calculated with the theoretical model of Eq. \eqref{eq:free_energy} for zero (black curves) and finite values (red and orange curves) of the symmetry-breaking field $\eta$. The panels correspond to different positions in the phase diagram of Fig.~\ref{fig:phase_diagram} on a path from a SVC ground state to an SSDW ground state ($w=0.2$, $-0.1\leq g\leq 0.5$). The insets display the SVDW order parameter, i.e., the order parameter of the vestigial phase associated with the SVC phase. Note that all calculations are performed in two dimensions so that there is no magnetic order. $r_0^*$ marks the transition into either the SVDW or nematic phase at $\eta=0$. (g)-(l) show the inverse of the nematic susceptibility. Note the change in the horizontal scale between panels with results for $g<0$ and panels with results for $g>0$.}
    \label{fig:theo}
\end{figure*}

We calculate the nematic susceptibility $\chi_{\textnormal{nem}}$ and the spin-vorticity density-wave (SVDW) order parameter $\zeta \equiv (\mathbf{M}_1 \times \mathbf{M}_2)\cdot \hat{z}$ via the one-loop self-consistent approximation in the SSDW and SVC regions of the phase diagram. We consider two-dimensional systems, hence magnetic order does not onset at any finite temperature due to Mermin-Wagner theorem. As a consequence, our results are only valid in the region prior to the onset of magnetic order. Coupling to the lattice degrees of freedom is implicitly incorporated in the Ginzburg-Landau parameter $g$, which is renormalized from its bare value \cite{Fernandes2012}. The full details of this calculation are described in Appendix~\ref{app:theory_details}. 

Fig.~\ref{fig:theo} shows the numerically evaluated nematic susceptibility on a path through the $g-w$ parameter space of Fig.~\ref{fig:phase_diagram}. Specifically, $g$ is changed from negative to positive while $w=0.2$ is kept constant, so that the ground state evolves from SVC to SSDW. Note that, $r_0^*$ denotes the transition temperature associated with the $\eta=0$ onset of the vestigial order, i.e. SVDW order for $g<0$ and nematic order for $g>0$, whereas there is not magnetic order at finite temperatures in our two-dimensional calculations. 
Starting with the case $\eta = 0$ (black curves in Fig. \ref{fig:theo}), the nematic susceptibility, $\chi_\textnormal{nem}$, remains finite for $g<0$, but does not diverge [panels (a)-(b) and (g)-(h)]. For $r_0<r_0^*$, when the system is in the SVDW phase, $\chi_\textnormal{nem}$ saturates at its $r_0^{*}$ value in our approximation, due to the changed ground state. 
The increase of $\chi_\textnormal{nem}$ on decreasing temperature indicates the proximity to the vestigial nematic phase associated with the SSDW state. On the other hand, $\chi_\textnormal{nem}$ diverges for $g>0$ if $\eta=0$ [panels (c)-(f) and (i)-(l)]. Note that for parameters well inside the SSDW region of the mean-field phase diagram, the calculated nematic susceptibility is almost perfectly Curie-Weiss-like within the temperature range plotted. The vestigial SVDW order parameter, shown in the insets, is completely absent for $g>0$ when $\eta=0$.

We consider now the case $\eta \neq 0$, shown by the orange and red curves in Fig. \ref{fig:theo}. Because $\eta$ is a conjugate field to the SVDW order, it always induces a finite SVDW order parameter. This results in the smearing of the SVDW transition in the $g<0$ region, where SVC is the $\eta = 0$ mean-field ground state [see insets in panels (a)-(b)], and in the triggering of a finite SVDW order parameter in the $g>0$ region, where SSDW is the $\eta = 0$ mean-field ground state [see insets in panels (c)-(f)]. Moreover, a finite $\eta$ generally suppresses the nematic susceptibility. For $g<0$ [panels (a)-(b) and (g)-(h)], the nematic susceptibility becomes even less divergent. For $g>0$ [panels (c)-(f) and (i)-(l)], the nematic susceptibility no longer diverges at $r_0=r_0^*$ and, depending on the size of $\eta$, may not diverge down to zero temperature. Furthermore, the temperature dependence of the nematic susceptibility deviates from a Curie-Weiss-like behavior near $r_0^*$. Importantly, these effects are only significant near the nematic transition temperature and for small enough $g$ [panels (c) and (i)], where the SVC and SSDW states are nearly degenerate in the mean-field phase diagram (roughly corresponding to the region between the solid and dashed lines in Fig.~\ref{fig:phase_diagram}). Away from this region, the temperature dependence of the nematic susceptibility is barely affected by a finite $\eta$ [see for instance panels (e)-(f) and (k)-(l)].

\section{Elastoresistance}\label{sec:elastoresistance}

\begin{figure}
    \centering
    \includegraphics[width=\columnwidth]{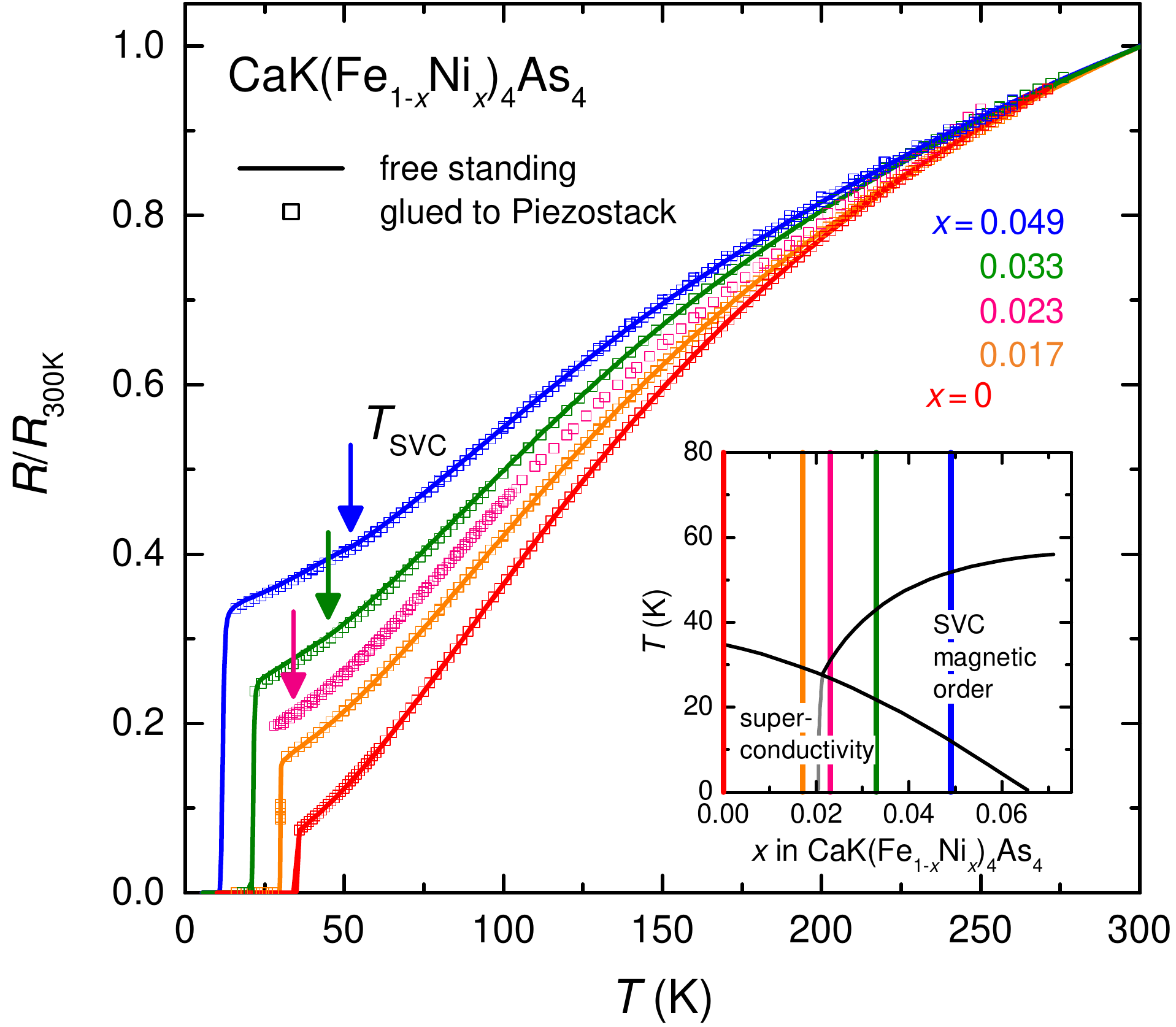}
    \caption{Electrical resistance of CaK(Fe$_{1-x}$Ni$_x$)$_4$As$_4$ normalized at 300 K of the five compositions studied here, comparing samples glued to a piezostack for elastoresistance measurements and samples in their freestanding state (before gluing). The resistivity is barely, or not measurably, changed. Data for free-standing samples have been published previously in Ref.~\onlinecite{Meier2018}. The transition into the magnetic SVC phase at $T_{\rm SVC}$ is marked by a subtle kink in the resistance data, which becomes obvious in the resistance derivative\cite{Meier2018}. $T_{\rm SVC}$ is first seen at $x=0.023$ and rises with increasing Ni substitution. The superconducting transition at $T_c=35$ K of CaKFe$_4$As$_4$ is gradually suppressed by Ni substitution. The inset shows the phase diagram adapted from Ref. \onlinecite{Meier2018} with the studied compositions indicated by vertical colored lines.}
    \label{fig:norm_resistance_measurements}
\end{figure}

\begin{figure}
    \centering
    \includegraphics[width=\columnwidth]{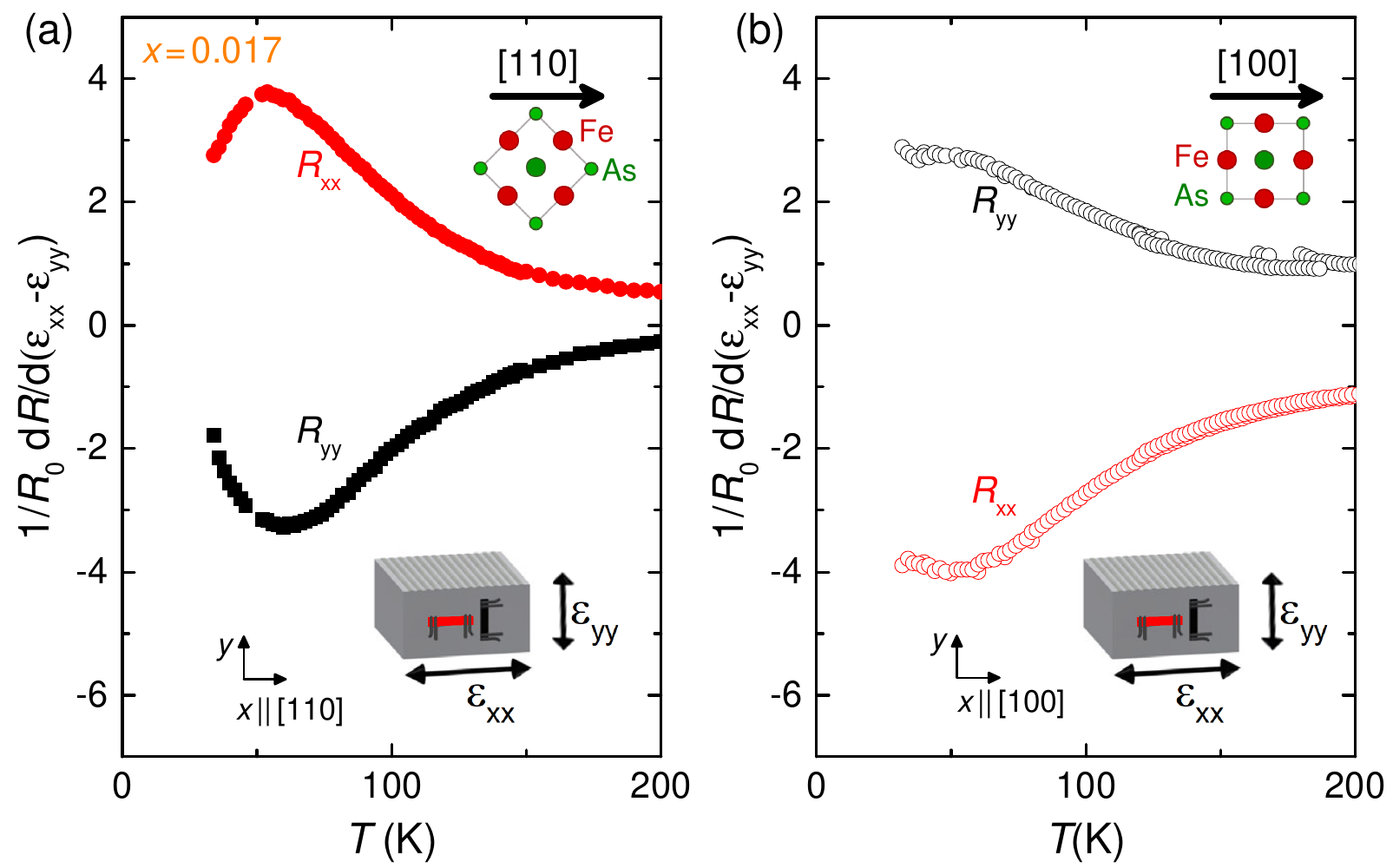}
    \caption{Elastoresistance of four samples cut from the same single crystal of CaK(Fe$_{0.983}$Ni$_{0.017}$)$_4$As$_4$. A pair of samples was cut along the [100] (a) and another along the [110] crystal direction (b), see insets in the upper right corner. In each pair, the two samples are oriented perpendicular to each other so that "longitudinal" and "transverse" elastoresistance along [100] and [110] are measured (depicted schematically in the lower right corners). The relative resistance change along one direction with respect to applied asymmetric strain $\varepsilon_{xx}-\varepsilon_{yy}$ is plotted. It is of similar magnitude along all directions. The sign change between perpendicular directions indicates that the $B_{1g}$ and $B_{2g}$ elastoresistance coefficients are dominant as compared to the isotropic $A_{1g}$ elastoresistance.}
    \label{fig:anistropic_resistance}
\end{figure}

Elastoresistance is by now a well-established probe of nematicity\cite{Chu2012,Kuo2013,Kuo2016, Palmstrom2017}. This is because the resistance anisotropy is a measure of the nematic order parameter, describing the loss of fourfold rotational symmetry, and anisotropic strain is the conjugate field to the nematic order parameter\cite{Chu2012}. Hence, the nematic susceptility $\chi_{\textnormal{nem}}$ is related to the elastoresistance,
\begin{equation}
    \chi_\textnormal{nem}=k \frac{1}{R_0}\frac{d\left(R_{xx}-R_{yy}\right)}{d\left(\varepsilon_{xx}-\varepsilon_{yy}\right)},
\end{equation}
with $x$ and $y$ the two directions rendered inequivalent by the formation of nematic order, $\varepsilon$ the strain components and $R_0$ an average resistance. Note that $\chi_\textnormal{nem}$ here corresponds to the bare nematic susceptibility unrenormalized by coupling to the crystal lattice, since this coupling is expected to be suppressed by the external application of strain\cite{Chu2012}. It reaches a finite threshold value at the nematic transition and tends to diverge at a bare nematic transition temperature $T_0$. The proportionality constant, $k$, describes the coupling between the elastoresistance and nematic fluctuations. It depends on details of the Fermi surface and electronic scattering and may, in principle, be doping and temperature dependent, so that the temperature dependence of the elastoresistance may not always be equated with the temperature dependence of $\chi_\mathrm{nem}$. 

\subsection{Elastoresistance results}

\begin{figure*}
    \centering
    \includegraphics[width=17.2cm]{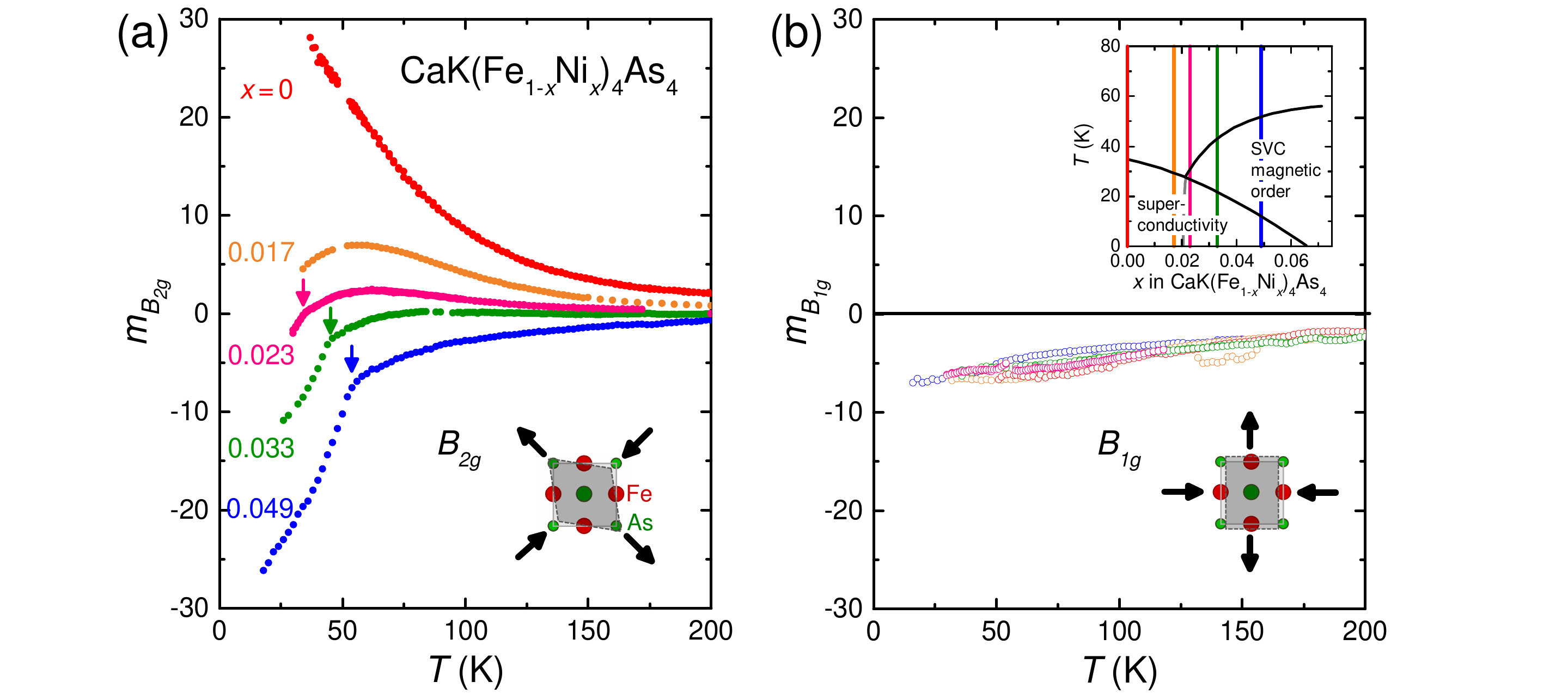}
    \caption{Elastoresistance coefficients (a) $m_{B_{2g}}$ and (b) $m_{B_{1g}}$ of CaK(Fe$_{1-x}$Ni$_
    {x}$)$_4$As$_4$. The lower insets show the corresponding lattice distortion schematically. A systematic doping dependence occurs only in the $B_{2g}$ channel with a sign change from negative to positive upon increasing Ni-substitution. The SVC transition, indicated by vertical arrows, is manifest as a clear kink. On the other hand, the elastoresistance in the $B_{1g}$ channel has no significant doping dependence and is only mildly temperature dependent. The data for pure CaKFe$_4$As$_4$ have been published previously in Ref. \onlinecite{Meier2016}. The upper inset in (b) shows the phase diagram adapted from Ref. \onlinecite{Meier2018} with the studied compositions indicated by colored vertical lines.}
    \label{fig:elastoresistance_coeffs}
\end{figure*}

Controlled strain is applied to samples via a piezoelectric stack to which they are glued. We first confirm that the samples are only negligibly affected by gluing to this substrate and that the elastoresistance reflects the properties of freestanding samples. Fig.~\ref{fig:norm_resistance_measurements} shows the resistance of the five studied compositions of CaK(Fe$_{1-x}$Ni$_x$)$_4$As$_4$. We directly compare the resistance of freestanding samples and samples that are glued to a piezostack for elastoresistance measurements. Samples of one composition were cut from a single $\sim2\times2$ mm$^2$ piece. 
The comparison shows that the resistance is barely --- and in many cases not even measurably --- affected by gluing the samples to the piezostack. 

In Fig. \ref{fig:anistropic_resistance} we show the resistance change induced by anisotropic strain $d(\Delta R/R)/d(\varepsilon_{xx}-\varepsilon_{yy})$ of the two pairs of samples cut from a crystal with $x=0.017$ as an example. The orientation of the samples with respect to the applied strain and the crystallographic directions are depicted in the lower right corner. Notably, the magnitude of the resistance change is similar for all four configurations and rather small, such that 1\% of distortion would induce less than 4\% resistance change. At the same time, the resistance change of two perpendicular samples is always of opposite sign.

We define our $x-y$ coordinate system such that $x$ is parallel to the poling direction of the piezostack. In the notation of the irreducible representations of the tetragonal $D_{4h}$ point group, samples aligned with their [100] (or [110]) axis along $x$ experience $B_{1g}$ (or $B_{2g}$) strain, given by the difference $\varepsilon_{xx}-\varepsilon_{yy}$. This induces a resistance change $(\Delta R/R_0)_{B_{1g}}$ [or $(\Delta R/R_0)_{B_{2g}}$, depending on the sample orientation] given by $(\Delta R/R)_{xx}-(\Delta R/R)_{yy}$ in both cases
\cite{Palmstrom2017}. In addition, there is also a finite isotropic strain component $\varepsilon_{A_{1g}}$, which induces an isotropic resistance change $(\Delta R/R_0)_{A_{1g}}=(\Delta R/R)_{xx}+(\Delta R/R)_{yy}$. The linear elastoresistance coefficients in the symmetry channel $\alpha$ are then defined as $m_\alpha=d(\Delta R/R_0)_{\alpha}/d\varepsilon_\alpha$.
Our data exhibit a clear sign change between the two perpendicular directions $R_{xx}$ and $R_{yy}$. We therefore find that the elastoresistance coefficients $m_{B_{1g}}$ and $m_{B_{2g}}$, which are related to the nematic susceptibilities in the $B_{1g}$ and $B_{2g}$ channels, respectively, dominate the response while $m_{A_{1g}}$ is rather small. We note that, with these conventions, the nematic susceptibility familiar from the 122-type compounds has been related to a divergent $B_{2g}$ elastoresistivity $\chi_\textnormal{nem}=k\,m_{B_{2g}}$.

Figures \ref{fig:elastoresistance_coeffs}(a) and (b) present the elastoresistance coefficients $m_{B_{1g}}$ and $m_{B_{2g}}$ of CaK(Fe$_{1-x}$Ni$_x$)$_4$As$_4$ over a wide range of the phase diagram. 
Clearly, only $m_{B_{2g}}$ has a remarkable temperature and doping dependence indicating the presence of sizeable nematic fluctuations. $m_{B_{2g}}$ of pure CaKFe$_4$As$_4$ is positive and increases on decreasing temperature, as noted previously\cite{Meier2016,Terashima2020}. Upon increasing Ni content, $m_{B_{2g}}$ decreases in magnitude and eventually changes sign, acquiring a complex temperature dependence. At the highest studied composition, $x=0.049$, $m_{B_{2g}}$ is negative in the whole temperature range and increases markedly in magnitude on decreasing temperature. The implications of this observed sign change are discussed below.
The SVC transition at $T_{\textnormal{SVC}}$ is marked by a clear kink and a further magnitude increase of $m_{B_{2g}}$ on decreasing temperature. This may suggest that  nematic fluctuations are sizeable even inside the SVC ground state but it could also be related to changes of the proportionality constant $k$. Note that this observation of $B_{2g}$-type nematic fluctuations in CaK(Fe$_{1-x}$Ni$_x$)$_4$As$_4$ is at odds with a recent Raman study, which reported the absence of any nematic fluctuations in CaKFe$_4$As$_4$ from an isotropic electronic Raman signal\cite{Zhang2018}.

In clear contrast to the behavior of $m_{B_{2g}}$, $m_{B_{1g}}$ is only mildly temperature dependent and has no measurable doping dependence. This demonstrates that $m_{B_{1g}}$ is not affected by the interactions that stabilize SVC magnetic order upon Ni substitution. We therefore find no evidence for $B_{1g}$ type nematicity in CaK(Fe$_{1-x}$Ni$_x$)$_4$As$_4$.

\subsection{Elastoresistance discussion}

\begin{figure}
    \centering
    \includegraphics[width=\columnwidth]{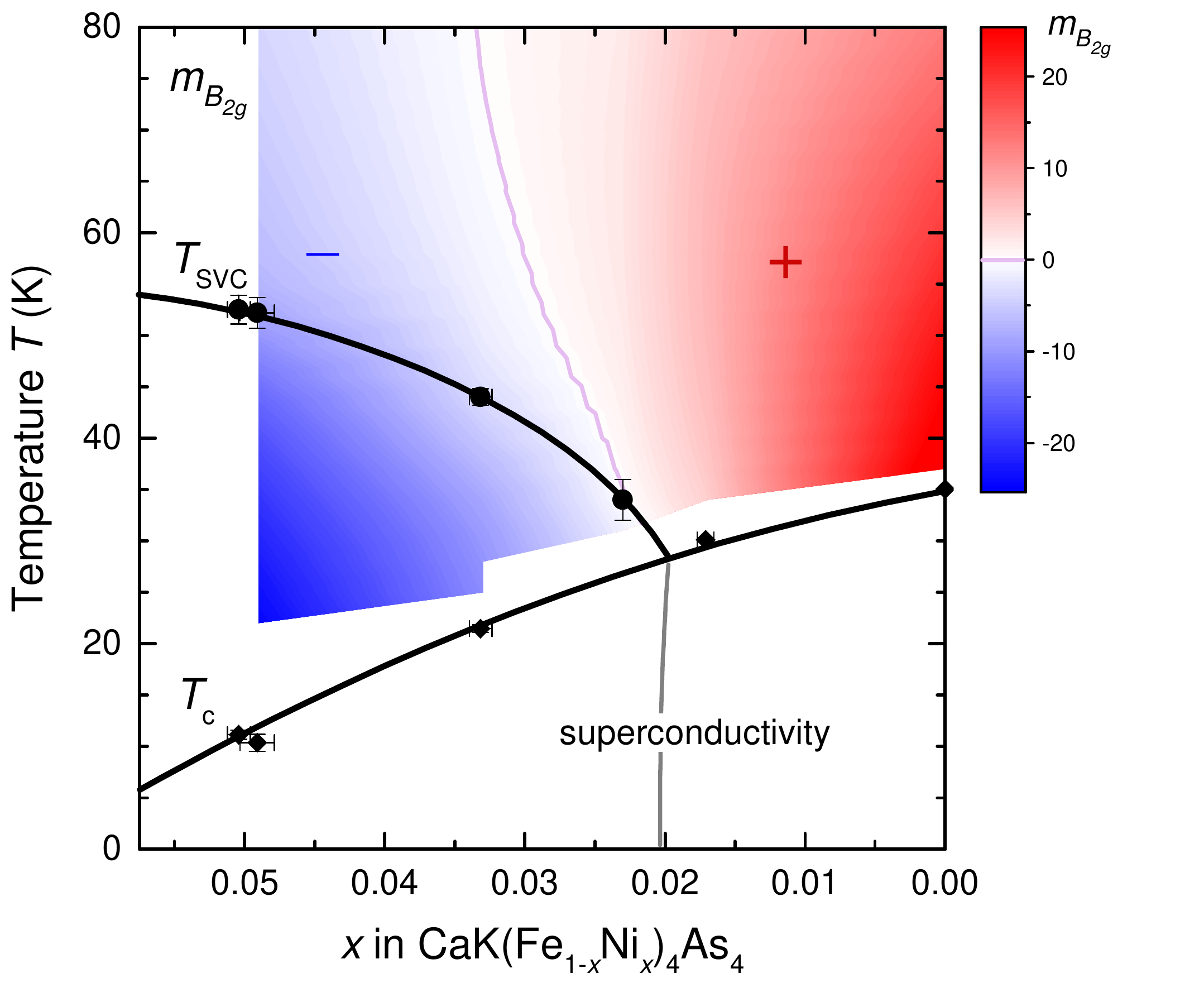}
    \caption{Value of the $B_{2g}$ elastoresistance coefficient as a color-coded map highlighting its sign change in the temperature-substitution phase diagram of CaK(Fe$_{1-x}$Ni$_x$)$_4$As$_4$. Note that the phase diagram is plotted for increasing ``hole-concentration" to facilitate comparison with hole-doped 122-type compounds such as Ba$_{1-x}$K$_x$Fe$_2$As$_2$\cite{Blomberg2013}.}
    \label{fig:B2g_elastoresistance_color_map}
\end{figure}

A prominent characteristic of the elastoresistance coefficient $m_{B_{2g}}$ of CaK(Fe$_{1-x}$Ni$_x$)$_4$As$_4$ is the sign change upon Ni substitution, as evident in Figs. \ref{fig:elastoresistance_coeffs}(a) and \ref{fig:B2g_elastoresistance_color_map}. Whereas $m_{B_{2g}}$ is proportional to the nematic susceptibility, the sign of $m_{B_{2g}}$ is given by the sign of the resistance anisotropy in the strain-induced orthorhombic state and eventually determined by the coupling constant $k$. We visualize the value of $m_{B_{2g}}$ in the substitution-temperature phase space of CaK(Fe$_{1-x}$Ni$_{x}$)$_4$As$_4$ in Fig. \ref{fig:B2g_elastoresistance_color_map} as a color-coded map. Note that the $x$-scale is inverted to reflect increasing hole-count in order to ease comparison with hole-doped 122 systems such as Ba$_{1-x}$K$_x$Fe$_2$As$_2$. 
The strain-induced resistance anisotropy is negative for low hole count (higher Ni content) and positive for higher hole count. Curiously, the sign change happens almost exactly at the same concentration where the SVC ordered phase emerges at low temperatures, which is around 2\% Ni content. 

Note that the doping evolution of $m_{B_{2g}}$ strongly differs from other iron-based systems such as Ba(Fe$_{1-x}$Co$_x$)$_2$As$_2$ or FeSe$_{1-x}$S$_x$, where $m_{B_{2g}}$ has a maximum in proximity of a putative quantum critical point\cite{Chu2012,Kuo2016,Hosoi2016,Sanchez2020}. However, there is a clear similarity with the previously reported sign change of the resistance anisotropy of stress-detwinned Ba$_{1-x}$K$_x$Fe$_2$As$_2$\cite{Blomberg2013}. In the latter case, the resistance anisotropy has been theoretically proposed to arise from the scattering off of anisotropic magnetic fluctuations in the iron-based materials\cite{Fernandes_Abrahams_Schmalian}, where the sign depends on the location of the scattering hotspots on the Fermi surface \cite{Blomberg2013, Tanatar2016}. 
The similarity with Ba$_{1-x}$K$_x$Fe$_2$As$_2$, and the dominant role of $m_{B_{2g}}$ with respect to $m_{B_{1g}}$, indicates that similar scattering from spin fluctuations plays an important role in determining the elastoresistance in CaK(Fe$_{1-x}$Ni$_x$)$_4$As$_4$.

\begin{figure}
    \centering
    \includegraphics[width=\columnwidth]{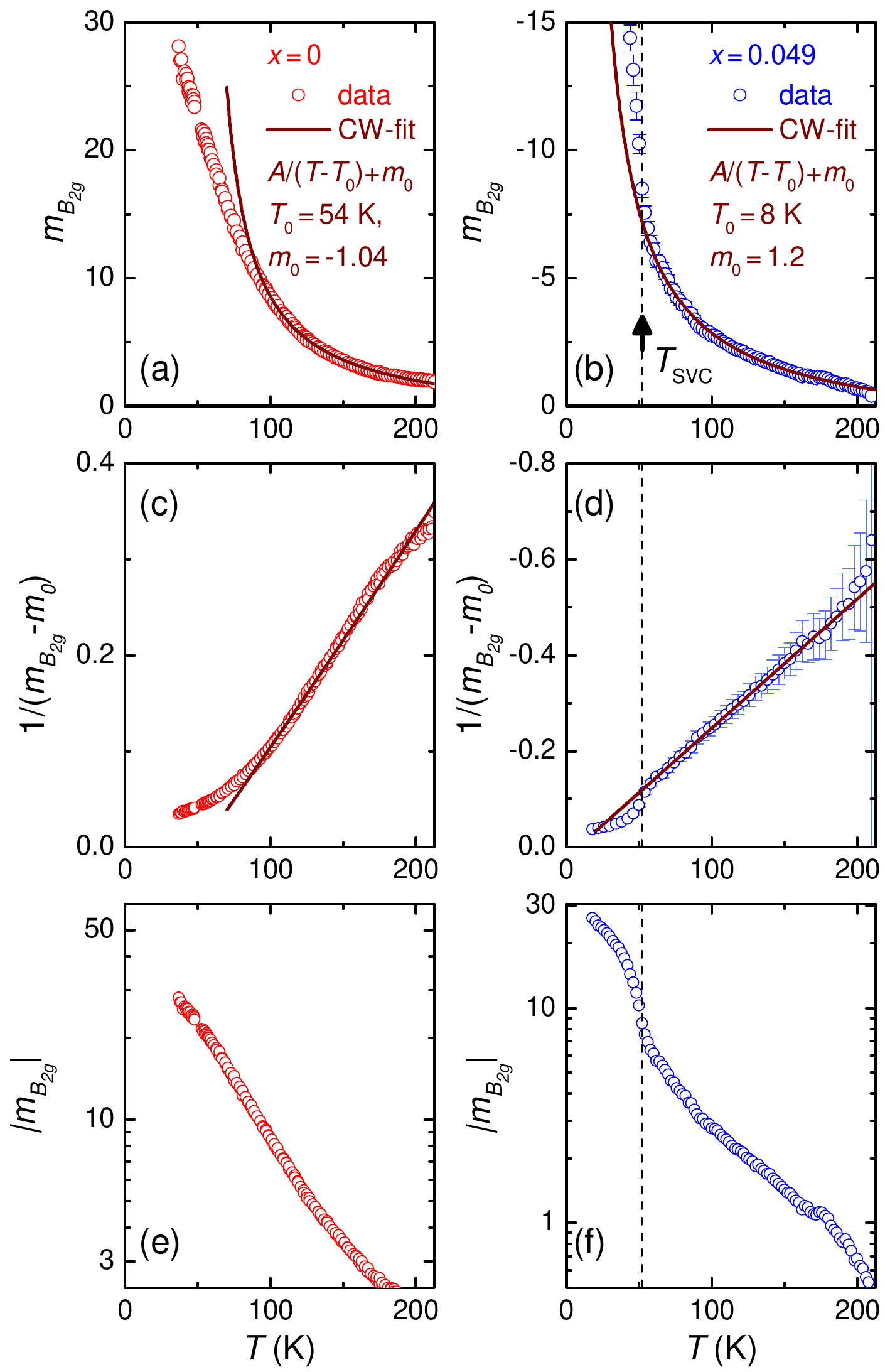}
    \caption{Analysis of the temperature dependence of the $B_{2g}$ elastoresistance coefficients of (a),(c),(e) pure CaKFe$_4$As$_4$ and (c),(d),(f) CaK(Fe$_{0.951}$Ni$_
    {0.049}$)$_4$As$_4$, i.e., the two compositions furthest away from the sign change of the coefficient. (a),(b) Elastoresistance coefficient $m_{B_{2g}}$ along with the results fitting to a Curie-Weiss law (CW-fit). (c),(d) Inverse of the data in (a),(b) highlighting deviations from the Curie-Weiss law. (e),(f) Data on a logarithmic vertical scale.  In CaK(Fe$_{0.951}$Ni$_{0.049}$)$_4$As$_4$ with $T_{\textnormal{SVC}}=52$ K, $m_{B_{2g}}$ follows a Curie-Weiss law well from $T_{\textnormal{SVC}}$ up to at least 200 K, beyond which the coefficient becomes too small to be accurately measured. In pure CaKFe$_4$As$_4$, $m_{B_{2g}}$ follows an approximate Curie-Weiss law at intermediate temperatures only. Below 150 K, its temperature dependence resembles an exponential decay.}
    \label{fig:B2g_temp_dependence}
\end{figure}

For some systems, it was previously argued that the proportionality constant $k$ is independent of temperature and therefore the temperature dependence of $m_{B_{2g}}$ directly reflects the temperature dependence of the $\chi_\textnormal{nem}$ \cite{Kuo2014,Sanchez2020}. CaK(Fe$_{1-x}$Ni$_x$)$_x$As$_4$ presents an example where this does not appear to be the case. Close to the composition-induced sign change, the temperature dependence of $m_{B_{2g}}$ of CaK(Fe$_{1-x}$Ni$_x$)$_x$As$_4$ is quite complex. Here, small details of the Fermi surface and scattering are likely temperature dependent and obscure the temperature dependence of the nematic susceptibility. In Fig.~\ref{fig:B2g_temp_dependence}, we therefore discuss the temperature dependence of $m_{B_{2g}}$ for the two compositions furthest from the sign change, i.e., $x=0$ and $x=0.049$, where $m_{B_{2g}}$ is likely dominated by the nematic susceptibility and not by $k$. Indeed, for 4.9\% Ni content, where SVC order onsets at $T_{\mathrm{SVC}}=52$ K, the temperature dependence of $m_{B_{2g}}$ is very well described by a Curie-Weiss law from $T_{\textnormal{SVC}}$ up to at least $200$ K with $m_{B_{2g}}\propto 1/(T-T_0)+m_0$, with $T_0=8$ K and a small constant contribution $m_0$ of the order of 1.  
Such a Curie-Weiss-like temperature dependence is expected from a mean-field type divergence of the nematic susceptibility, with a bare nematic transition at $T_0$ and where the actual nematic transition would occur at a higher temperature. In the present case, however, such a transition is presumably preempted by the onset of the SVC phase.

The temperature dependence of $m_{B_{2g}}$ of pure CaKFe$_4$As$_4$ is less simple. It can be approximately described by a Curie-Weiss type increase at intermediate temperatures, but deviations are significant both at high and at low temperatures. We therefore show in Fig. \ref{fig:B2g_temp_dependence}(e) the absolute value of $m_{B_{2g}}$ on a log-scale with the data for CaK(Fe$_{0.951}$Ni$_{0.049}$)$_4$As$_4$ shown similarly in panel (f) for comparison. $m_{B_{2g}}$ of pure CaKFe$_4$As$_4$ approximately follows an exponential decay function below $\sim150$ K. The origin of such a behavior is unclear, though a similar temperature dependence has been seen in optimally doped Ba$_{0.6}$K$_{0.4}$Fe$_2$As$_2$ with similar carrier concentration\cite{Meier2016}. In Ba$_{0.6}$K$_{0.4}$Fe$_2$As$_2$, the close proximity between a SSDW state and the tetragonal CSDW state\cite{Boehmer2015II} may lead to deviations from a simple Curie-Weiss law. Concerning CaKFe$_4$As$_4$, it is conceivable that, also for this material, the proximity of competing magnetic ground states leads to a non-Curie-Weiss-like temperature dependence of the nematic susceptibility. This is in line with the results of our theoretical modeling shown in Fig.~\ref{fig:theo}, which find that the nematic susceptibility deviates most from Curie-Weiss behavior near the boundary between SSDW and SVC phase.

\section{Young's modulus}\label{sec:elastic_modulus}

To shed further light on the nematic susceptibility of CaK(Fe$_{1-x}$Ni$_x$)$_4$As$_4$, we turn to the elastic modulus. This complementary thermodynamic quantity allows us to sidestep the complexities of the prefactor of elastoresistance. In a nematic system, the nematic order parameter couples bilinearly to an orthorhombic distortion. This means that the corresponding shear modulus is renormalized by the nematic susceptibility $\chi_{\textnormal{nem}}$. The higher $\chi_{\textnormal{nem}}$, the softer the shear modulus becomes. For the $B_{2g}$-type nematic susceptibility found here, the relevant elastic modulus is $C_{66}$, given by 
\begin{equation}
C_{66}=C_{66,0}-\lambda^2\chi_{\textnormal{nem}}\,,\label{eq:C66}    
\end{equation}
where $C_{66,0}$ is the bare shear modulus, and  $\chi_{\textnormal{nem}}$ is the bare nematic susceptibility (as is also relevant for the elastoresistance experiments) and $\lambda$ is a coupling constant\cite{Fernandes2010,Boehmer2014,Boehmer2016}. The bending modulus of thin samples is proportional to the elastic Young's modulus. In particular, Young's modulus along the [110] direction is given by
\begin{equation}
    Y_{[110]}=4\left(\frac{1}{C_{66}}+\frac{1}{\gamma}\right)^{-1}
    \label{eq:Y110}
\end{equation}
with 
\begin{equation}
    \gamma=\frac{C_{11}+C_{12}}{2}-\frac{C_{13}^2}{C_{33}}\,.
\end{equation}
Only $C_{66}$ is expected to soften across a $B_{2g}$-type nematic transition and the combination of elastic constants $\gamma$ is not expected to have a strong temperature dependence. Hence, the shear modulus $C_{66}$ dominates the temperature dependence of $Y_{[110]}$. For example, a Curie-Weiss-like temperature dependence $\chi_\textnormal{nem}=A/(T-T_{0})$, leads to a softening of the shear modulus according to
\begin{equation}
    C_{66}=C_{66,0}\frac{T-T_s}{T-T_0}\textnormal{ with } T_s=T_0+\frac{\lambda^2A}{C_{66,0}}.
\end{equation}
Assuming $\gamma$ to be temperature independent, Young's modulus will follow a similar temperature dependence with modified characteristic temperature $T_0'$ (Ref. \onlinecite{Massat2016})
\begin{equation}
    Y_{[110]}\propto\frac{T-T_s}{T-T_{0}'}\textnormal{ with }  T_0'=\frac{T_0+\alpha T_s}{1+\alpha},\label{eq:Y110Tdep}
\end{equation}
where the parameter $\alpha=\frac{C_{66,0}}{\gamma}$.

\subsection{Young's modulus results}

\begin{figure}
    \centering
    \includegraphics[width=\columnwidth]{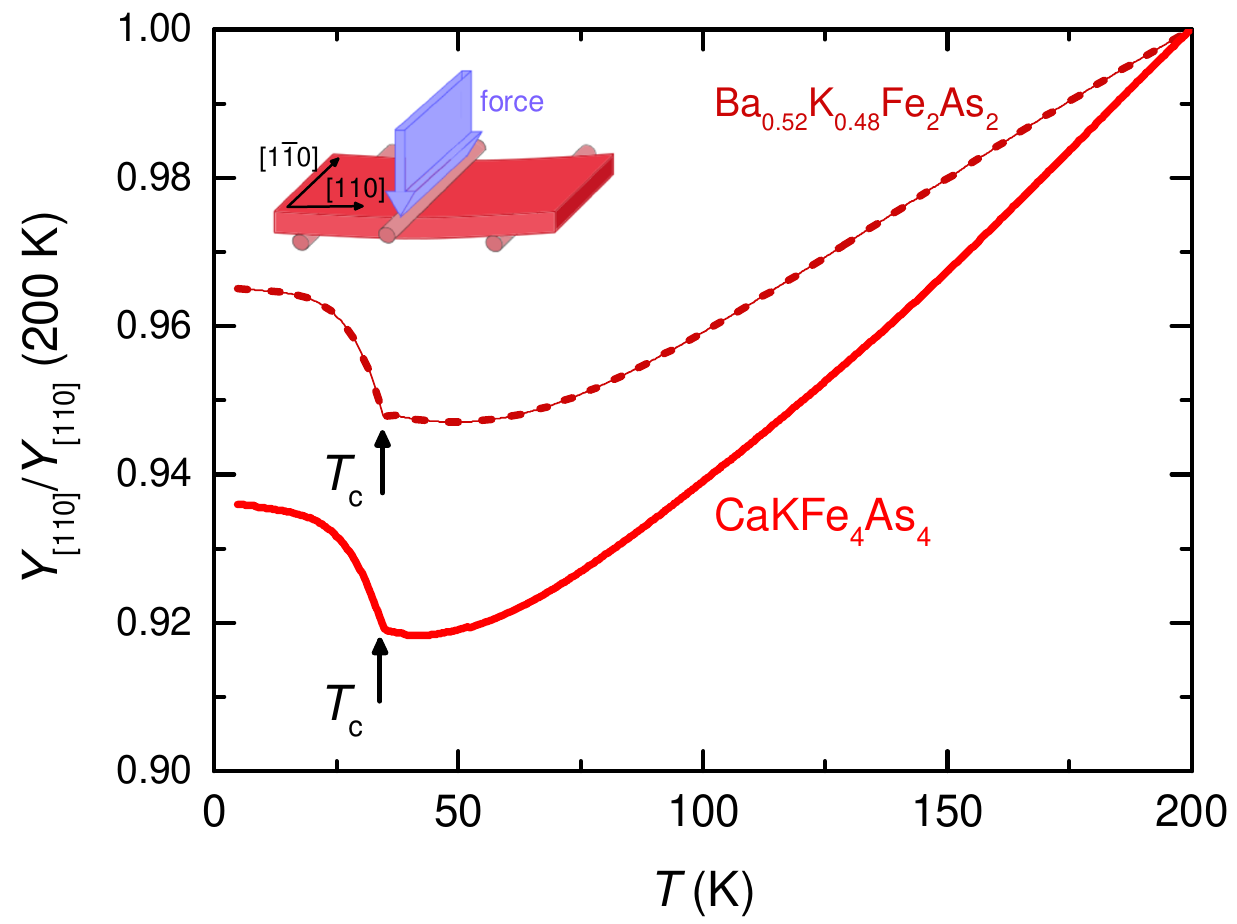}
    \caption{Young's modulus $Y_{[110]}$ of undoped CaKFe$_4$As$_4$ measured using a 3-point bending technique in a high-resolution capacitance dilatometer\cite{Boehmer2014} (see inset). There is a clear softening on decreasing temperature and a distinct hardening below $T_c$. Overall, the temperature dependence is very similar to Ba$_{0.52}$K$_{0.48}$Fe$_2$As$_2$ with a practically identical charge count.}
    \label{fig:young_modulus_undoped}
\end{figure}

Young's modulus $Y_{[110]}$ of the two extreme compositions, pure CaKFe$_4$As$_4$ and 5\% Ni-doped CaK(Fe$_{0.095}$Ni$_{0.05}$)$_4$As$_4$ were studied using samples from other batches which contained relatively large single crystals. $Y_{[110]}$ of pure CaKFe$_4$As$_4$ is presented in Fig. \ref{fig:young_modulus_undoped}. Consistent with nematic fluctuations in the $B_{2g}$ channel, a marked softening on decreasing temperature is observed. $Y_{[110]}$  decreases almost linearly on decreasing temperature at higher temperatures and flattens below $\sim75$ K. Below $T_c=35$ K, $Y_{[110]}$ hardens by $\sim2\%$. This behaviour closely resembles Ba$_{0.52}$K$_{0.48}$Fe$_2$As$_2$ with practically the same charge count, also shown in Fig. \ref{fig:young_modulus_undoped}. The hardening below $T_c$ is even identical in magnitude for the two materials. The similarity suggest that they both host similar nematic fluctuations interacting with superconductivity. 
The softening of $Y_{[110]}$ indicates an increased $B_{2g}$ nematic susceptibility. However, the observed temperature dependence cannot be described by Eq.~\eqref{eq:Y110Tdep}, implying that the nematic susceptibility does not follow a Curie-Weiss law. 
The nematic susceptibility of CaKFe$_4$As$_4$ as inferred from Young's modulus is shown in Fig. \ref{fig:nematic_susc_from_young} and discussed below. 

\begin{figure}
    \centering
    \includegraphics[width=\columnwidth]{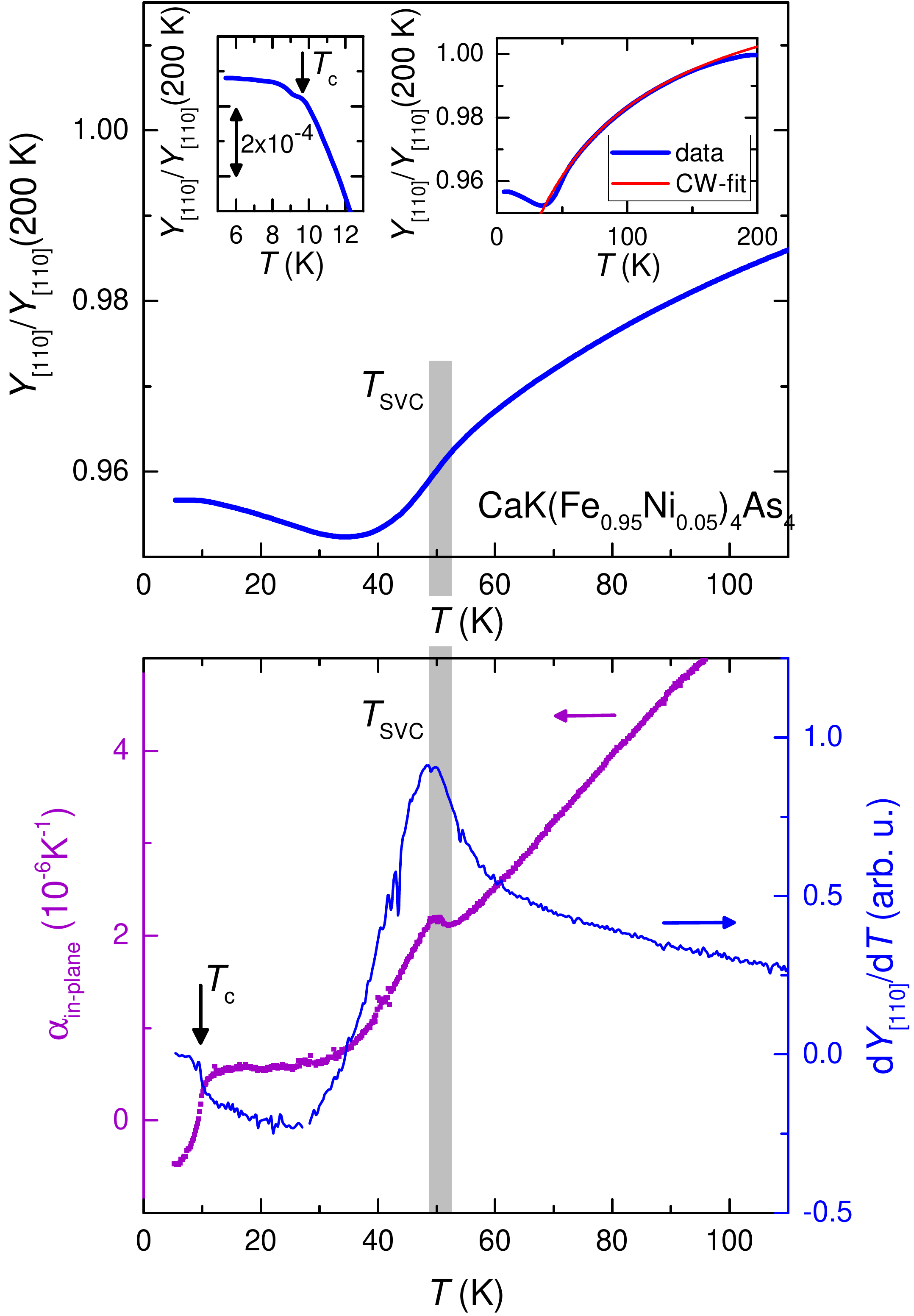}
    \caption{(a) Young's modulus $Y_{[110]}$ of CaK(Fe$_{0.95}$Ni$_{0.05}$)$_4$As$_4$ with an SVC-type magnetic transition at $\sim51$ K, manifesting in a subtle anomaly. The left inset shows a magnified view close to $T_c$, the right inset an expanded view up to 200 K. The red line is a fit of the data to Eq.~\eqref{eq:Y110Tdep}. (b) Left axis: In-plane thermal expansion coefficient $\alpha_{\textnormal{in-plane}}=1/L(dL/dT)$ ($L$ is the sample length along [110]) of the same sample. $T_{\textnormal{SVC}}$ and $T_c$ are well marked by their step-like anomalies. Right scale: Temperature derivative of $Y_{[110]}$ from panel (a), revealing the corresponding anomalies in the elastic modulus. $T_{\textnormal{SVC}}$ is marked by an inflection point of $Y_{[110]}$, or a maximum of its derivative.}
    \label{fig:young_modulus_doped}
\end{figure}

$Y_{[110]}$ of CaK(Fe$_{0.095}$Ni$_{0.05}$)$_4$As$_4$ is presented in Fig. \ref{fig:young_modulus_doped}. At this composition, $Y_{[110]}$ softens on decreasing temperature, following almost perfectly the temperature dependence described in Eq.~\eqref{eq:Y110Tdep} with $T_s=-58$ K and $T_0'=-66$ K. 
The effect of the SVC transition on the elastic modulus is rather subtle. This transition has previously been shown to be second order with a mean-field-type step-like anomaly in the specific heat\cite{Meier2018}. To locate $T_{\rm SVC}$ precisely for our particular sample, we measured its coefficient of uniaxial thermal expansion $\alpha_{\textnormal{in-plane}}=1/L(dL/dT)$, where $L$ is the sample length along [110] (Fig. \ref{fig:young_modulus_doped}). $\alpha_{\textnormal{in-plane}}$ exhibits a clear jump of $0.3\times10^{-6}$ K$^{-1}$ centered at $T_{\rm SVC}=51$ K with a width of 4 K, indicated by the grey bar. Such a shape is expected for the second-order transition\cite{Meier2018}. The finite width likely arises from small inhomogeneities in Ni-content and internal stresses. 

$Y_{[110]}$ has an inflection point centered at $49$ K with a width of 4 K, as revealed by the derivative $dY_{[110]}/dT$ in Fig. \ref{fig:young_modulus_doped}(b). This temperature is slightly lower than $T_{\rm SVC}$ deduced from $\alpha_\textnormal{in-plane}$. It is possible that the large bending stress along [110], to which the sample is subjected during the measurement and which favors a nematic distortion, slightly suppresses the SVC transition. An inflection point of the elastic modulus is a curious anomaly at a second-order phase transition. Note that the crystal unit cell does not change at $T_{\rm SVC}$\cite{Meier2018}, hence, no elastic constant is expected to show critical softening on approaching $T_{\rm SVC}$. Instead, a small step-like anomaly similar to the anomaly in $\alpha_{\textnormal{in-plane}}$ would be expected in the Young's modulus. Thermodynamic relations even allow us to predict its size to be $<0.01\%$ of the high-temperature value\cite{notejumpY110_1144}, but this tiny anomaly is not resolved in the current data. 
Interestingly, an overall similar behavior with an inflection point at the magnetic transition has been seen in Young's modulus of Ba$_{0.638}$Na$_{0.326}$Fe$_2$As$_2$\cite{Wang2018}, a compound that has recently been proposed to show SVC-type magnetic order as well\cite{Sheveleva2020}. Furthermore, an inflection point in the nematic susceptibility is indeed expected in the presence of a finite $\eta$ from our theoretical modelling, see the discussion below.
The continuing softening of the bending modulus indicates an increase of the nematic susceptibility on decreasing temperature even inside the SVC magnetic phase. 

$Y_{[110]}$ eventually starts to harden on cooling. The hardening is arrested below the superconducting transition, resulting in a tiny kink. Overall, the response of the elastic modulus to superconductivity is much smaller than in pure CaKFe$_4$As$_4$ or Ba$_{0.638}$Na$_{0.326}$Fe$_2$As$_2$\cite{Wang2018}.

\subsection{Young's modulus discussion}

To extract the nematic susceptibility from the Young's modulus data we solve Eqs.~\eqref{eq:C66} and \eqref{eq:Y110} for the nematic susceptibility,
\begin{equation}
    \frac{\lambda^2\chi_\textnormal{nem}}{C_{66,0}}=1-\left(4\frac{C_{66,0}}{Y_{[110]}}-\frac{C_{66,0}}{\gamma}\right)^{-1}\,.
\end{equation}
Note that $\lambda^2\chi_\textnormal{nem}/C_{66,0}$ is dimensionless and reaches a value of $1$ at the nematic transition \cite{Boehmer2016}. To disentangle the different contributions, the elastic constants of CaKFe$_4$As$_4$ have been determined from first principles using density functional theory. For this, different deformations of the initial tetragonal structure have been imposed and the resulting stress tensor has been calculated, see Appendix \ref{app:DFT_details}. We find for the bare elastic constants, unrenormalized by coupling to nematic fluctuations,
\begin{equation}
    C_{ij,0}=\begin{pmatrix}
    133 & 15  & 44 & 0   & 0   & 0\\
    15  & 133 & 44 & 0   & 0   & 0\\
    44  & 44  & 55 & 0   & 0   & 0\\
    0    & 0    & 0   & 15 & 0   & 0\\
    0    & 0    & 0   & 0   & 15 & 0\\
    0    & 0    & 0   & 0   & 0   & 19\\
    \end{pmatrix}
    \,\text{GPa}
\end{equation}
and $C_{66,0}/\gamma=0.49$. Finally, the absolute value of $Y_{[110]}$ could not be experimentally measured with sufficient precision. We therefore normalize $Y_{[110]}$ at 200 K. The resulting additional parameter $C_{66,0}/Y_{[110]}(200\mathrm{ K})$ has been chosen based on a comparison to the 122-type systems \cite{Boehmer2014} and on the fit shown in Fig. \ref{fig:young_modulus_doped}(a). Varying this parameter affects the shape of the inferred $\chi_\textnormal{nem}$ only mildly.

\begin{figure}
    \centering
    \includegraphics[width=\columnwidth]{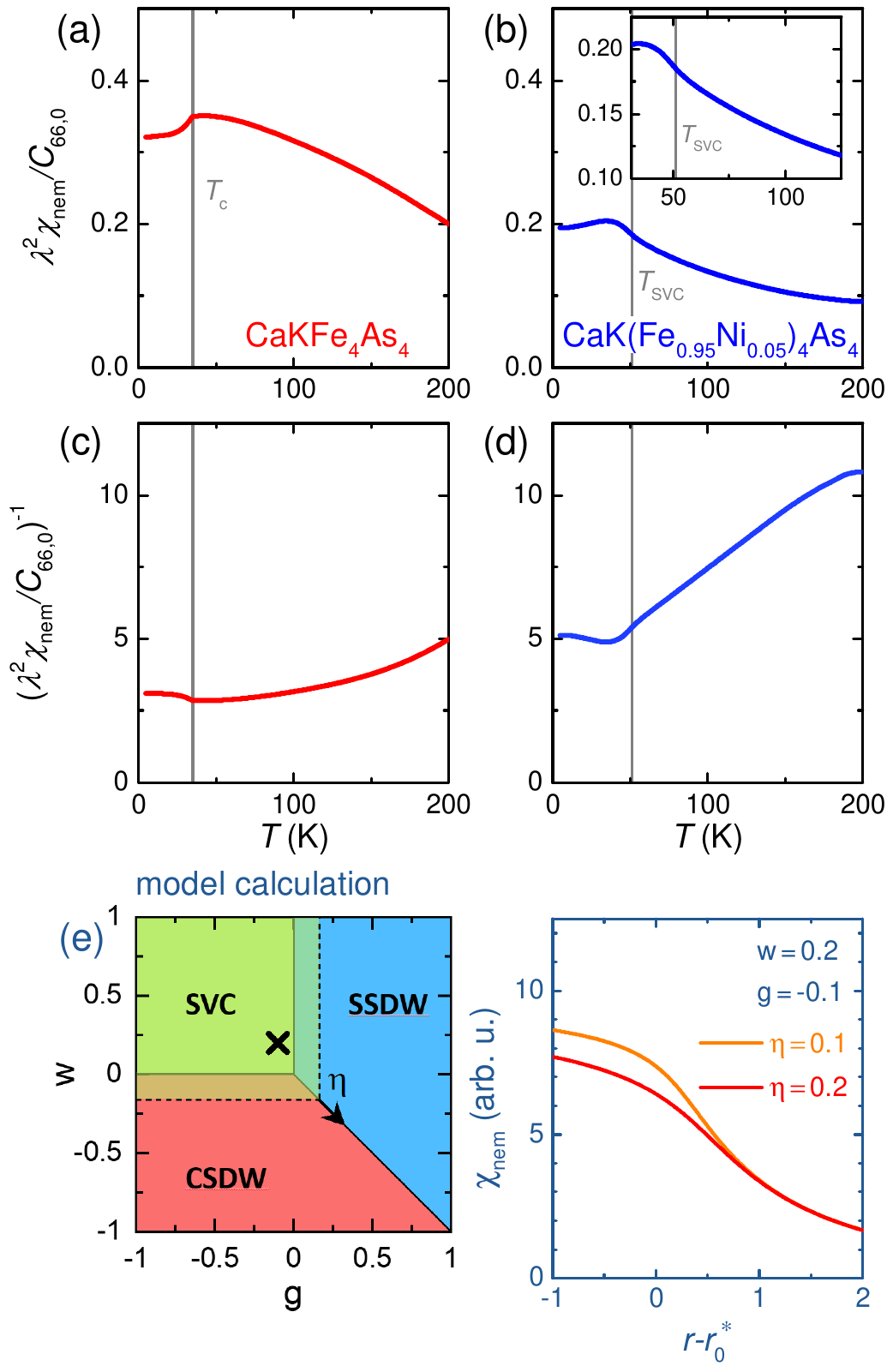}
    \caption{Nematic susceptibility inferred from Young's modulus data. (a), (b) Dimensionless nematic susceptibility, in which $\lambda^2\chi_{\textnormal{nem}}/C_{66,0}=1$ at the nematic transition. The parameter $C_{66,0}/Y_{[110]}(\textnormal{200 K})=0.815$ (corresponding to $\lambda^2\chi_\textnormal{nem}/C_{66,0}=0.2$) has been chosen in (a) based on analogies with the 122 system; the choice $C_{66,0}/Y_{[110]}(\textnormal{200 K})=0.778$ (corresponding to $\lambda^2\chi_\textnormal{nem}/C_{66,0}=0.0925$) in (b) is based on the Curie-Weiss fit. The inset in (b) shows the same data on an expanded scale. (c), (d) Inverse of the nematic susceptibility from panels (a), (b). (e) Position in the $g-w$ parameter space of Fig. \ref{fig:phase_diagram} (marked by a cross) and theoretical nematic susceptibility [duplicated from Fig. \ref{fig:theo}(g)] calculated at this position on an expanded scale.}
    \label{fig:nematic_susc_from_young}
\end{figure}

Fig.~\ref{fig:nematic_susc_from_young} presents the obtained dimensionless nematic susceptibility of CaKFe$_4$As$_4$ and CaK(Fe$_{0.095}$Ni$_{0.05}$)$_4$As$_4$. $\chi_\textnormal{nem}$ of pure CaKFe$_4$As$_4$ has a sub-Curie-Weiss-like temperature dependence [Fig. \ref{fig:nematic_susc_from_young}(a),(c)]. This is in qualitative agreement with the elastoresistance data. However, the saturation of the Young's modulus at lower temperatures is not seen in the elastoresistance data and the detailed temperature dependence remains unexplained.

The nematic susceptibility of CaK(Fe$_{0.095}$Ni$_{0.05}$)$_4$As$_4$ indeed shows an enhancement on decreasing temperature following a Curie-Weiss law [Fig. \ref{fig:nematic_susc_from_young}(b),(d)]. This is consistent with the temperature dependence of the elastoresistance coefficient $m_{B_{2g}}$. However, the characteristic temperatures are lower. Fitting with Eq.~\eqref{eq:Y110Tdep} and using $\alpha=C_{66,0}/\gamma=0.49$, a Weiss temperature $T_0=-70$ K and a $T_s=-58$ K are obtained. $T_0$ is much smaller than expected from the temperature dependence of $m_{B_{2g}}$. A similar discrepancy between elastic modulus and elastoresistance has been noted previously for Co-substituted BaFe$_2$As$_2$ and FeSe\cite{Boehmer2016}, though it is especially pronounced in the present system. We note that experimental uncertainties arising from the small sample size might affect $T_0$ deduced from Young's modulus, even though the overall temperature dependence is robust.

We now compare our experimental results with the theoretical calculations for the nematic susceptibility presented in Sec. \ref{sec:nemsus_theo}. As shown in Fig. \ref{fig:theo}, the main effect of the effective field $\eta$ generated by the inequivalent As positions is to suppress the nematic susceptibility and modify its temperature dependence. This effect is only significant at temperatures comparable to the onset of nematic/SVDW order and near $g=0$, i.e. near the mean-field phase boundary between SSDW and SVC in Fig. \ref{fig:phase_diagram}. Importantly, in this regime, the nematic susceptibility is generally enhanced as temperature is lowered, regardless of the whether the mean-field ground state is SSDW or SVC. Thus, our experimental results on CaKFe$_4$As$_4$ are consistent with a scenario in which the system is near the SVC-SSDW degeneracy point.

An interesting question is whether the hypothetical system with $\eta \rightarrow 0$ would have an SSDW or an SVC ground state -- in other words, whether the SVC phase in CaK(Fe$_{1-x}$Ni$_x$)$_4$As$_4$ is only stabilized by the inequivalent As positions. The theoretical nematic susceptibility curves shown in Fig. \ref{fig:theo} display deviations from Curie-Weiss behavior for both $g>0$ and $g<0$ regimes. Thus, the experimentally observed Curie-Weiss deviations for the $x=0$ sample could in principle be consistent with either scenario. It is tempting, nevertheless, to note that the theoretical nematic susceptibility in the $g<0$ regime shows an inflection point around the SVDW transition temperature for $\eta \neq 0$, as shown in Fig. \ref{fig:nematic_susc_from_young}(e). Interestingly, the experimentally extracted nematic susceptibility of the $x=0.05$ sample also shows an inflection point, but near the SVC transition temperature (see inset in Fig. \ref{fig:nematic_susc_from_young}(a)). While our theoretical model does not describe the SVC phase, because it is a strictly two-dimensional model, it is known that the SVC and SVDW phases may have very close transition temperatures in anisotropic three-dimensional systems \cite{Fernandes2016}. Of course, the limitations of our analysis imply that further studies are required to establish whether the inflection point observed here can really be attributed to an intrinsic SVDW instability that would happen even if $\eta \rightarrow 0$.

 \section{Summary and conclusions}\label{sec:conclusions}
 We presented an extensive study of the evolution of nematic fluctuations in Ni-doped CaKFe$_4$As$_4$ using elastoresistance and elastic modulus measurements. We find that the $B_{2g}$ nematic susceptibility is significant despite the absence of either nematic or SSDW order, whereas nematic fluctuations are absent in the $B_{1g}$ symmetry channel.
 
 The elastoresistance measurements feature a remarkable sign change as a function of doping. We interpret this as a sign that the coefficient between $m_{B_{2g}}$ and the nematic susceptibility $\chi_{\rm nem}$ is not a constant and even dominates the behavior of the elastoresistance over a certain doping and temperature range close to this sign change. As an alternative thermodynamic probe of the nematic susceptibility, we present Young's modulus measurements supported by first-principles estimates of the elastic constants. These are qualitatively consistent with the elastoresistance measurements and the combination of experimental methods forms a reliable basis for comparison with theory. 
 
 At high Ni content, we find a Curie-Weiss-like increase of the nematic susceptibility on decreasing temperature towards the magnetic SVC transition at $T_{\rm SVC}$. However, the nematic Curie-Weiss temperature is rather low and both probes reveal mean-field-type behavior of nematic fluctuations that are far from being critical at $T_{\rm SVC}$. The behavior is in contrast to the pure compound CaKFe$_4$As$_4$, for which the nematic susceptibility clearly deviates from a Curie-Weiss law and increases more mildly on decreasing temperature. The two experimental probes do not agree on the precise temperature dependence, but both of these results are similar to previous data in the hole-doped Ba$_{1-x}$K$_x$Fe$_2$As$_2$ system obtained by the respective probes.
 
 To understand these findings we carried out a comprehensive theoretical study of the nematic susceptibility in the case where the FeAs-layer possesses two inequivalent As sites, as realized in CaK(Fe$_{1-x}$Ni$_x$)$_4$As$_4$. The structural peculiarity is modeled by an effective field $\eta$ that couples to the SVDW order, the vestigial phase of the spin-vortex crystal magnetic phase. 
 The theoretical results are generally consistent with the experimental observations of an inflection point of $\chi_{\rm nem}$ near the SVC transition temperature and of deviations from Curie-Weiss-like behavior that indicate a suppression of nematic fluctuations. Both effects are caused by a small $\eta$ field, when the system is intrinsically near the degeneracy point between SSDW and SVC magnetic orderings. Such enhanced magnetic fluctuations between nearly-degenerate states may play an important role in enhancing pairing tendencies.

\begin{acknowledgments}
The project was partially funded by the Deutsche Forschungsgemeinschaft (DFG, German Research Foundation) - TRR 288 - 422213477. A.E.B and P.W.W. acknowledge support by the Helmholtz Association under contract number VH-NG-1242. Work at Ames Laboratory (A.E.B, W.R.M., M.X., G.D., S.L.B., P.C.C.) was supported by the U.S. Department of Energy, Office of Basic Energy Science, Division of Materials Sciences and Engineering and was performed at the Ames Laboratory. Ames Laboratory is operated for the U.S. Department of Energy by Iowa State University under Contract No. DE-AC02-07CH11358. W.R.M. and G.D. were supported in part by the Gordon and Betty Moore Foundation’s EPiQS Initiative through Grant GBMF4411. The contribution from M. M. was supported by the Karlsruhe Nano Micro Facility (KNMF). Theoretical modeling (F.C., M.H.C, and R.M.F) was supported by the U.S. Department of Energy, Office of Science, Basic Energy Sciences, Materials Science and Engineering Division, under Award No. DE-SC0020045. V.B. would like to thank the Goethe University Frankfurt for providing the computational resources which were used for calculating the elastic constants using density functional theory.
\end{acknowledgments}

\appendix
\section{Theoretical modeling}\label{app:theory_details}

We consider two-dimensional systems in what follows. Hence, magnetic order will not onset at any finite temperature.

\subsection{Nematic susceptibility for $\eta=0$}

We start by reviewing the derivation of the nematic susceptibility in the case $\eta=0$, as previously discussed elsewhere (see for instance Ref. \onlinecite{Fernandes2010}). Our starting point is the action~\cite{Fernandes2012II}
\begin{align}
	\mathcal{S} = \int_q \chi_{\rm mag}^{-1} \left( \mathbf{M}_1^2 + \mathbf{M}_2^2 \right)  &+ \frac{u}{2}\int_x \left( \mathbf{M}_1^2 + \mathbf{M}_2^2 \right)^2 \nonumber \\ &- \frac{g}{2} \int_x \left( \mathbf{M}_1^2 - \mathbf{M}_2^2 \right)^2\,,\label{eq:action_no_eta}
\end{align}
where $\int_q \equiv T \int \frac{\mathrm{d}^d q}{(2\pi)^d}$. Here $\chi_{\rm mag}^{-1}$ is the momentum-dependent inverse magnetic propagator. Following Ref. \onlinecite{Fernandes2012II}, we perform a Hubbard-Stratonovich decoupling and introduce two fields, $\psi$ and $\phi$, corresponding to Gaussian and nematic fluctuations, respectively. The resulting action reads
\begin{align}
	\mathcal{S} = -\int_q \frac{\psi^2}{2u} + \int_q\frac{\phi^2}{2g} &+ \int_q (\chi_{\rm mag}^{-1} + \psi)  \left( \mathbf{M}_1^2 + \mathbf{M}_2^2 \right) \nonumber \\ &+ \phi\int_q \left( \mathbf{M}_1^2 - \mathbf{M}_2^2\right)\,.
\end{align}
Introducing a conjugate field to the spin-nematic order, $- h \left( \mathbf{M}_1^2 - \mathbf{M}_2^2 \right)$ and integrating out the magnetic degrees of freedom we find, in the paramagnetic state:
\begin{equation}
	\mathcal{S}_{\rm eff} = -\int_q \frac{\psi^2}{2u} + \int_q\frac{\phi^2}{2g} + \int_q \log \left[ \left(\chi_{\rm mag}^{-1} + \psi \right)^2 - (\phi + h)^2 \right]\,,
\end{equation}
where we specialized to the case $N=2$ (here $N$ denotes the number of components of the magnetic order parameters $\mathbf{M}_1$ and $\mathbf{M}_2$) to facilitate direct comparison with the subsequent section. Rescaling $\phi \rightarrow \phi - h$, we can calculate the nematic susceptibility as
\begin{equation}
	\chi_{\rm nem} = \frac{\partial^2 \log\mathcal{Z}}{\partial h^2}\bigg|_{h\rightarrow 0}\,,
\end{equation}
where $\mathcal{Z}$ is the partition function, $\mathcal{Z} = \int \mathcal{D}[\mathbf{M}_1, \mathbf{M}_2, \psi, \phi]e^{-\mathcal{S}_{\rm eff}}$. We find
\begin{equation}
	\chi_{\rm nem} = \frac{\int_q \widetilde{\chi}_{\rm mag}^2}{1-g\int_q \widetilde{\chi}_{\rm mag}^2}\,,
\end{equation}
where $\widetilde{\chi}_{\rm mag} = \left(\chi^{-1}_{\rm mag} + \psi \right)^{-1}$. This can be further simplified by evaluating the self-consistent equations:
\begin{eqnarray}
	\frac{\partial \mathcal{S}}{\partial \psi} = 0 \Rightarrow \psi = u \int_q \frac{2(\chi_{\rm mag}^{-1} +\psi)}{(\chi_{\rm mag}^{-1} + \psi)^2 - \phi^2} \\
	\frac{\partial \mathcal{S}}{\partial \phi} = 0 \Rightarrow \phi = g\int_q \frac{2\phi}{(\chi_{\rm mag}^{-1} + \psi)^2 - \phi^2} \,.
\end{eqnarray}

To proceed we make a number of simplifying assumptions. We let $\chi_{\rm mag}^{-1} = r_0 +q^2$ where $q$ is measured relative to the ordering vector. We then define $r \equiv r_0 + \psi$ and the (inverse) nematic susceptibility simply reads
\begin{equation}
	\chi_{\rm nem}^{-1} = \frac{2\pi}{T_{\rm mag,0}}\left( r - \bar{g} \right)\,,
\end{equation}
with $\bar{g}=\frac{gT_{\rm mag,0}}{2\pi}$. The above self-consistent equations become
\begin{align}
	r &= r_0 + u \int_q \frac{2(r +q^2)}{(r+q^2)^2-\phi^2} \\
	\phi &= g \int_q \frac{2\phi}{(r+q^2)^2-\phi^2}\,.
\end{align}
In the disordered non-nematic phase $\phi=0$, and $r$ can be obtained from
\begin{equation}
	r = r_0 + u \int_q \frac{2}{r+q^2}\,.
\end{equation}
In two dimensions, this integral diverges in the ultraviolet, and we introduce a high-momentum cutoff $\Lambda$:
\begin{equation}
	r = r_0 + \frac{u T_{\rm mag,0}}{2\pi} \int_0^{\Lambda} \frac{2q\mathrm{d}q}{r+q^2}\,,
\end{equation}
Defining $\bar{u} \equiv \frac{uT_{\rm mag,0}}{2\pi}$ and $\bar{r}_0 \equiv r_0 + 2\bar{u}\log \Lambda$ we find the simple equation of motion
\begin{equation}
	r = \bar{r}_0 - \bar{u} \log r\,. \label{eqBB}
\end{equation}
This permits us to evaluate the nematic susceptibility for $\eta = 0$ and $g>0$. The factor of $T_{\rm mag,0}$ sets the overall scale of the mean-field magnetic transition.

\subsection{Nematic susceptibility for $\eta \neq 0$}

To compute the nematic susceptibility for $g<0$ and to account for the role of the conjugate field $\eta$, we include the terms~\cite{Fernandes2016,Meier2018,Christensen2019}
\begin{eqnarray}
	-\eta \int_q (\mathbf{M}_1 \times \mathbf{M}_2) \cdot \hat{z} + 2w \int_x \left( \mathbf{M}_1 \cdot \mathbf{M}_2 \right)^2
\end{eqnarray}
in the action of Eq.~\ref{eq:action_no_eta}. A third Hubbard-Stratonovich field is now required to decouple the action, yielding 
\begin{align}
	\mathcal{S} &= -\int_q \frac{\psi^2}{2(u+w)} + \int_q \frac{\phi^2}{2(g+w)} + \int_q \frac{\zeta^2}{2w} \nonumber \\ & + \int_q \left(\chi_{\rm mag}^{-1} +\psi \right) \left( \mathbf{M}_1^2 + \mathbf{M}_2^2 \right) + \int_q \phi \left(\mathbf{M}_1^2 - \mathbf{M}_2^2 \right) \nonumber \\ & -2 \int_q \left( \zeta + \frac{\eta}{2} \right) \left( \mathbf{M}_1 \times \mathbf{M}_2 \right) \cdot \hat{z}\,.
\end{align}
To obtain this expression it is necessary to rewrite the ($\mathbf{M}_1 \cdot \mathbf{M}_2$) term in terms of a cross-product. Integrating out the magnetic fluctuations, we find
\begin{align}
	\mathcal{S}_{\rm eff} &= -\int_q \frac{\psi^2}{2(u+w)} + \int_q \frac{\phi^2}{2(g+w)} + \int_q \frac{\zeta^2}{2w} \nonumber \\ & + \int_q \log\left[ (\chi_{\rm mag}^{-1} +\psi)^2 -\phi^2 - \left(\zeta + \frac{\eta}{2} \right)^2 \right]\,.
\end{align}
With the assumptions made above and by letting $\zeta + \frac{\eta}{2} \rightarrow \zeta$ the self-consistent equations become
\begin{eqnarray}
	r &=& r_0 + (u+w)\int_q \frac{2(r+q^2)}{(r+q^2)^2- \zeta^2 - \phi^2} \label{eq:eom_with_eta_1} \\
	\phi &=& (g+w)\int_q\frac{2\phi}{(r+q^2)^2- \zeta^2 - \phi^2} \\
	\zeta &=& w \int_q \frac{2\zeta}{(r+ q^2)^2- \zeta^2 - \phi^2} + \frac{\eta}{2}\,. \label{eq:eom_with_eta_3}
\end{eqnarray}
These equations demonstrate that the vestigial order of the SVC order, the SVDW order parameter $\zeta$, is always finite if $\eta$ is finite. The expression for the nematic susceptibility can be obtained in a fashion similar to the above, and we find
\begin{equation}
	\chi_{\rm nem} = \left(\frac{T_{\rm mag,0}}{2\pi} \right)\frac{\frac{\arctanh \left[\frac{\zeta}{r} \right]}{\zeta}}{1- (\bar{g}+\bar{w}) \frac{\arctanh \left[\frac{\zeta}{r} \right]}{\zeta}}.
\end{equation}
The factor of $T_{\rm mag,0}$ arises from evaluation of the integrals in Eqs.~\eqref{eq:eom_with_eta_1}--\eqref{eq:eom_with_eta_3}, as the case of the previous subsection. As previously, this simply sets the overall temperature scale of the transition. The evaluation of the nematic susceptibility requires solving the self-consistent equations to find how $r$ and $\zeta$ behave as functions of $\eta$. Focusing on the non-nematic phase, $\phi=0$, we obtain
\begin{eqnarray}
	r &=& \zeta \coth \left[ \frac{\zeta - \frac{\eta}{2}}{\bar{w}} \right] \\
	\bar{r}_0 &=& \zeta \coth\left[ \frac{\zeta - \frac{\eta}{2}}{\bar{w}} \right] + (\bar{u}+ \bar{w}) \log \left[\frac{\zeta}{\sinh\left[\frac{\zeta -\frac{\eta}{2}}{\bar{w}} \right]} \right] \label{eqAA}
\end{eqnarray}
where, as before, $\bar{u} = \frac{uT_{\rm mag,0}}{2\pi}$ and similarly for $w$, and $\bar{r}_0 = r_0 + 2(\bar{u}+\bar{w})\log \Lambda $. These equations are evaluated numerically, and the results are shown in Fig.~\ref{fig:theo}. The parameters used there were $\bar{u}=1.0$ and $\bar{w}=0.2$, while $\eta$ and $\bar{g}$ were varied. Note that, for $g<0$, the location of the transition $\bar{r}_0^*$ is given by $\bar{r}_0^*= \bar{w} + (\bar{w}+\bar{u})\ln \bar{w}$ whereas, for $g>0$, $\bar{r}_0^* = \bar{g} + \bar{u} \ln \bar{g}$. Note that the coefficients referred to in the main text are the barred constants, $\bar{u}$, $\bar{g}$, and $\bar{w}$.

\section{\textit{Ab initio} elastic constants
\label{app:DFT_details}}

First-principles calculation of the elastic constants is based on the following expression for the deformation energy of the tetragonal CaKFe$_4$As$_4$:
\begin{equation}
  \begin{array}{rl}
	F = &\frac12 C_{xxxx} (\varepsilon_{xx}^2 + \varepsilon_{yy}^2) + \frac12 C_{zzzz} \varepsilon_{zz}^2 + \\[5pt]
	&C_{xxzz} (\varepsilon_{xx} \varepsilon_{zz} + \varepsilon_{yy} \varepsilon_{zz} ) + C_{xxyy} \varepsilon_{xx} \varepsilon_{yy} +\\[5pt]
	&2C_{xyxy} \varepsilon_{xy}^2 + 2 C_{xzxz} (\varepsilon_{xz}^2 + \varepsilon_{yz}^2)
  \end{array}  \label{e:deformation_energy}
\end{equation}

The strain tensor $\varepsilon_{ij}$ is defined in the usual way:
\begin{equation}
	\varepsilon_{ij} = \frac12 \left( \frac{\partial \varepsilon_i}{\partial x_j} + \frac{\partial \varepsilon_j}{\partial x_i} \right), \hspace{5pt} i,j=x,y,z \label{e:strain_tensor}
\end{equation}

Different components of the stress tensor are obtained from (\ref{e:deformation_energy}) as follows:
\begin{equation}
  \sigma_{ij} = -\frac{\partial F}{\partial \varepsilon_{ij}}, \hspace{5pt} i,j=x,y,z \label{e:stress_tensor}
\end{equation}

In order to estimate the elastic constants in the expansion (\ref{e:deformation_energy}) for CaKFe$_4$As$_4$, we use density functional theory to calculate the stress tensor for different kinds of lattice deformations of the initial tetragonal structure (see below). The calculations are based on the projector-augmented wave method as available in Vienna \textit{Ab initio} Simulation Package (VASP). The exchange-correlation energy is described by the PBE parameterization of the generalized-gradient approximation. For the wavefunctions, we choose the energy cutoff of 800~eV and, for the integration within the Brillouin zone, the $\Gamma$-centered $(5\times 5\times 5)$ $k$-mesh. For a given lattice deformation, the internal atomic positions are optimized until the ionic forces become sufficiently small. Following the conclusions of the previous studies,\cite{Kaluarachchi2017,Borisov2018} we impose the spin-vortex configurations on the Fe sublattice, in order to take into account the spin fluctuations to a first approximation and to obtain reliable structural parameters.

The following types of deformations were considered:
\begin{enumerate}
	\item[a)] The unit cell is deformed along the [100] direction leading to the lattice vectors
	\begin{equation}
		\begin{array}{rlcr}
			\vec{a}_1 &= ( 1 + \delta,& 0,& 0) \\[5pt]
			\vec{a}_2 &= (0,& 1,& 0) \\[5pt]
			\vec{a}_3 &= (0,& 0,& c/a)
		\end{array}
	\end{equation}

	and the strain tensor with the only non-zero component $\varepsilon_{xx} = \delta \ll 1$. Based on (\ref{e:deformation_energy}) and (\ref{e:stress_tensor}), the resulting stress tensor can be calculated:
	\begin{equation}
		\sigma = \left(
		\begin{array}{ccc}
			C_{xxxx} & 0 & 0 \\[5pt]
			0 & C_{xxyy} & 0 \\[5pt]
			0 & 0 & C_{xxzz}
		\end{array}
		\right) \cdot \delta
	\end{equation}

	\item[b)] The unit cell is deformed along the [001] direction leading to the lattice vectors
	\begin{equation}
		\begin{array}{rlcr}
			\vec{a}_1 &= (1,& 0,& 0) \\[5pt]
			\vec{a}_2 &= (0,& 1,& 0) \\[5pt]
			\vec{a}_3 &= (0,& 0,& c/a\cdot(1 + \delta))
		\end{array}
	\end{equation}

	and the strain tensor with the only non-zero component $\varepsilon_{zz} = \delta \ll 1$. The stress tensor reads:
	\begin{equation}
		\sigma = \left(
		\begin{array}{ccc}
			C_{xxzz} & 0 & 0 \\[5pt]
			0 & C_{xxzz} & 0 \\[5pt]
			0 & 0 & C_{zzzz}
		\end{array}
		\right) \cdot \delta
	\end{equation}

	\item[c)] The unit cell is skewed in the $xy$-plane leading to the lattice vectors
	\begin{equation}
		\begin{array}{rlcr}
			\vec{a}_1 &= (1,& \delta,& 0) \\[5pt]
			\vec{a}_2 &= (0,& 1,& 0) \\[5pt]
			\vec{a}_3 &= (0,& 0,& c/a)
		\end{array}
	\end{equation}

	and the strain tensor with the only non-zero component $\varepsilon_{xy} = \delta/2 \ll 1$. The stress tensor reads:
	\begin{equation}
		\sigma = \left(
		\begin{array}{ccc}
			0 & 4 C_{xyxy} & 0 \\[5pt]
			4 C_{xyxy} & 0 & 0 \\[5pt]
			0 & 0 & 0
		\end{array}
		\right) \cdot \frac{\delta}{2}
	\end{equation}

	\item[d)] The unit cell is skewed in the $xz$-plane leading to the lattice vectors
	\begin{equation}
		\begin{array}{rlcr}
			\vec{a}_1 &= (1,& 0,& \delta) \\[5pt]
			\vec{a}_2 &= (0,& 1,& 0) \\[5pt]
			\vec{a}_3 &= (0,& 0,& c/a)
		\end{array}
	\end{equation}

	and the strain tensor with the only non-zero component $\varepsilon_{xz} = \delta/2 \ll 1$. The stress tensor reads:
	\begin{equation}
		\sigma = \left(
		\begin{array}{ccc}
			0 & 0 & 4 C_{xzxz} \\[5pt]
			0 & 0 & 0 \\[5pt]
			4 C_{xzxz} & 0 & 0
		\end{array}
		\right) \cdot \frac{\delta}{2}
	\end{equation}
\end{enumerate}

Calculation of the stress tensor in the aforementioned cases for different small values of $\delta > 0$ using density functional theory, as described above, allowed to determine separately all the elastic constants:
\begin{equation}
  \begin{array}{lr}
      C_{xxxx} = 132.6\:\text{GPa} & C_{zzzz} = 54.7\:\text{GPa} \\[5pt]
      C_{xxyy} = 14.6\:\text{GPa} & C_{xxzz} = 44.1\:\text{GPa} \\[5pt]
      C_{xyxy} = 18.9\:\text{GPa} & C_{xzxz} = 14.5\:\text{GPa}
  \end{array}
\end{equation}

Following the conventional definition $1,2,3,4,5,6=xx,yy,zz,yz,xz,xy$, these values can be summarized in the matrix form:
\begin{equation}
    C_0=\begin{pmatrix}
    13.3 & 1.5  & 4.4 & 0   & 0   & 0\\
    1.5  & 13.3 & 4.4 & 0   & 0   & 0\\
    4.4  & 4.4  & 5.5 & 0   & 0   & 0\\
    0    & 0    & 0   & 1.5 & 0   & 0\\
    0    & 0    & 0   & 0   & 1.5 & 0\\
    0    & 0    & 0   & 0   & 0   & 1.9\\
    \end{pmatrix}
    \cdot 10^{10}\,\text{Pa}
\end{equation}
\\
\newpage
\bibliography{referencespnictides}

\end{document}